\documentclass[aps,showpacs,showkeys]{revtex4}

\usepackage{amsmath,amssymb}
\usepackage{epsfig}
\begin{document}

\title{\textbf{Similarity solutions of the Fokker-Planck equation with time-dependent coefficients}}
\author{W.-T. Lin and C.-L. Ho}%\\
%\address
\affiliation{Department of Physics, Tamkang University\\
Tamsui 25137, Taiwan, R.O.C.}

\date{Oct 25, 2011} %{Jun 16, 2011}

%\maketitle  % for LaTex

\begin{abstract}

In this work, we consider the solvability of the Fokker-Planck
equation with both time-dependent drift and diffusion coefficients
by means of the similarity method. By the introduction of the
similarity variable, the Fokker-Planck equation is reduced to an
ordinary differential equation.  Adopting the natural requirement
that the probability current density vanishes at the boundary, the
resulting ordinary differential equation turns out to be
integrable, and the probability density function can be given in
closed form. New examples of exactly solvable Fokker-Planck
equations are presented, and their properties analyzed.
\end{abstract}



\maketitle

%\newpage

\section{Introduction}

The Fokker-Planck equation (FPE) is one of the basic tools which
is widely used for studying the effect of fluctuations in
macroscopic systems \cite{RIS:1996}. It has been employed in many
areas: physics, chemistry, hydrology, biology, even finance, and
others. Because of its broad applicability, it is therefore of
great interest to obtain solutions of the FPE for various physical
situations.  Many methods, including analytical, approximate and
numerical ones, have been developed to solve the FPE.

Generally, it is not easy to find analytic solutions of the FPE.
Exact analytical solutions of the FPE are known for only a few
cases, such as linear drift and constant diffusion coefficients
\cite{RIS:1996}. In most cases, one can only solve the equation
approximately, or numerically. The well-known analytic methods for
solving the FPE include a change of variables, perturbation
expansion, eigenfunction expansion, variational approach, Green's
function, path integral, moment method, and the continued-fraction
method (for a review of these methods, see eg. \cite{RIS:1996}).
Symmetry methods have also been quite useful in solving the FPE
\cite{BC:1974}.  Two interesting analytic approximation approaches
are the WKB analysis \cite{RIS:1996} and the normal mode analysis
\cite{CCR:1981}. The finite-difference \cite{DM:1996} and
finite-element methods \cite{ZTZ:2005} are among some useful
numerical methods. Most of these methods, however, are concerned
only with FPEs with time-independent diffusion and drift
coefficients.

Solving the FPEs with time-dependent drift and/or diffusion
coefficient is in general an even more difficult task. It is
therefore not surprising that the number of papers on such kind of
FPE is far less than that on the FPE with time-independent
coefficients.  Some recent works on the FPE with time-dependent
diffusion coefficients appear in
\cite{GMNT:2005,KSF:2005,GNT:2009}, and works involving
time-dependent drift coefficients can be found in
\cite{LM:2000,HY:2008,LH:2011}. Refs.
\cite{WH:1980,OK:1985,SS:1999} consider FPEs with both
time-dependent diffusion and drift coefficients. The symmetry
properties of the one-dimensional FPE with arbitrary coefficients
of drift and diffusion are investigated in \cite{SS:1999}. Such
properties may in some cases allow one to transform the FPE into
one with constant coefficients. It is proved that symmetry group
of these equations can be of one, two, four or six parameters and
the criteria are also obtained.

In our previous work \cite{LH:2011}, we have considered, within
the framework of a perturbative approach, the similarity solutions
of a class of FPEs  which have constant diffusion coefficients and
small time-dependent drift coefficients.  Motivated by our
previous work, in this paper we would like to study the
solvability of the FPE with both time-dependent drift and
diffusion coefficients by means of the similarity method.

One advantage of the similarity method is that it allows one to
reduce the order of a partial differential equation
\cite{ZIL:1992}.  Thus the FPE can be transformed into an
ordinary differential equation which may be easier to solve.
However, for the FPE to admit similarity solutions, it must
possess proper scaling property under certain scaling
transformation of the basic variables. This all boils down to the
scaling behaviors of the drift and diffusion coefficients. To
study these scaling behaviors is the main aim of the present work.

The plan of the paper is as follows.  Sect.~II discusses the
scaling form of the FPE.  Then in Sect.~III the similarity method
is applied to reduce the FPE into an ordinary differential
equation, and the exact solution of the probability density
expressed in closed form. Some new examples of exactly solvable
FPEs with similarity solutions are presented in Sect.~IV and V.
Sect.~VI concludes the paper.

\section{Scaling of Fokker-Planck equation}

The general form of the FPE in $(1+1)$-dimension is
\begin{align}\label{E2.1}
\frac{\partial W(x,t)}{\partial t}=\Big[-\frac{\partial}{\partial x}
D^{(1)}(x,t)+\frac{\partial^2}{\partial x^2}D^{(2)}(x,t)\Big]W(x,t)\;,
\end{align}
where $W(x,t)$ is the probability distribution function,
$D^{(1)}(x,t)$ is the drift coefficient and $D^{(2)}(x,t)$ the
diffusion coefficient. The drift coefficient represents the
external force acting on the particle, while the diffusion
coefficient accounts for the effect of fluctuation. $W(x,t)$ as a
probability distribution function should be normalized, $i.e.$,
$\int_{\textstyle\mbox{\small{domain}}}W(x,t)\,dx=1$ for $t\geq
0$. The domains we shall consider in this paper are the real line
$x\in (-\infty,\infty)$, and the half lines $x\in [0,\infty)$ and
$(-\infty,0]$.

As mentioned in the Introduction, it is difficult, if not
impossible, to find exact solutions of the general FPE with
time-dependent drift and diffusion coefficients. Here we shall be
content with a more modest aim by seeking a special class of
solutions, namely, the similarity solutions of the FPE.  Such
solutions are possible, provided that the FPE possesses certain
scaling symmetry. Below we shall show that, if both the drift and
diffusion coefficients assume proper scaling forms, then the FPE
can be solved with the similarity method.

Consider the scale transformation
\begin{align}\label{E2.2}
&\bar{x}=\varepsilon^a x\;\;\;,\;\;\;\bar{t}=\varepsilon^b t,
\end{align}
where $\varepsilon$, $a$ and $b$ are real parameters. Suppose
under this transformation, the probability density function and
the two coefficients scale as
\begin{align}\label{E2.3}
\bar{W}(\bar{x},\bar{t})=\varepsilon^c W(x,t),~~
\bar{D}^{(1)}(\bar{x},\bar{t})=\varepsilon^d
D^{(1)}(x,t)\;\;\;,\;\;\;\bar{D}^{(2)}(\bar{x},\bar{t})=\varepsilon^e
D^{(2)}(x,t).
\end{align}
Here $c$, $d$ and $e$ are also some real parameters.  Written in
the transformed variables, eq.(\ref{E2.1}) becomes
\begin{align}\label{E2.4}
\varepsilon^{b-c}\frac{\partial \bar{W}}{\partial
\bar{t}}&=\Big[-\varepsilon^{a-c-d}\frac{\partial}{\partial
\bar{x}}\bar{D}^{(1)}(\bar{x},\bar{t})+\varepsilon^{2a-c-e}\frac{\partial^2}{\partial
\bar{x}^2}\bar{D}^{(2)}(\bar{x},\bar{t})\Big]\bar{W}(\bar{x},\bar{t})\;.
\end{align}
One sees that if the scaling indices satisfy $b=a-d=2a-e$, then
eq.(\ref{E2.4}) has the same functional form as eq.(\ref{E2.1}).
In this case, the FPE admits similarity solutions.   We shall
present such solutions below.

\section{Similarity method}

The similarity method is a very useful method for solving a
partial differential equation which possesses proper scaling
behavior. One advantage of the similarity method is to reduce the
order of a partial differential equation through some new
independent variables (called similarity variables), which are
certain combinations of the old independent variables such that
they are scaling invariant, i.e., no appearance of parameter
$\varepsilon$, as a scaling transformation is performed.

In our case, the second order FPE can  be transformed into an
ordinary differential equation which may be easier to solve. Here
there is only one similarity variable $z$, which can be defined as
\begin{align}\label{E3.1}
z\equiv\frac{x}{t^{\alpha}}\;,\;\;\;\mbox{where}\;\;\;
\alpha=\frac{a}{b}\;\;\;\mbox{and}\;\;\;a\;,b\neq 0\;.
\end{align}
For $a\;,b\neq 0$, one has $\alpha\neq 0\;,\infty$. In what
follows, we derive the scaling forms and closed form solutions of
the probability and current density functions.

\subsection{Probability distribution function}

The general scaling form of the probability density function
$W(x,t)$ is $W(x,t)=(t^{\delta_1}/x^{\lambda_1})y(z)$, where
$\lambda_1$ and $\delta_1$ are two real parameters and $y(z)$ is a
scale-invariant function under the same transformation
(\ref{E2.2}). From the assumed scaling behavior of $W(x,t)$ in
(\ref{E2.3}), we have $-c/a=\lambda_1-(\delta_1/\alpha)$. Without
loss of generality and for clarity of presentation, we hereby
assume the parameters $(\lambda_1,\delta_1)=(0,\alpha c/a)$. This
gives
\begin{align}\label{E3.2}
W(x,t)=t^{\alpha\frac{c}{a}}y(z)\;.
\end{align}
The normalization of the distribution function is
\begin{align}\label{E3.5}
\int_{\mbox{\small{domain}}}\,W(x,t)\,dx=\int_{\mbox{\small{domain}}}\,
\Big[t^{\alpha(1+\frac{c}{a})}\,y(z)\Big]\,dz=1\;.
\end{align}
For the above relation to hold at all $t\geq 0$, the power of $t$
should vanish, and so one must have $c=-a$, and thus
\begin{equation}
W(x,t)=t^{-\alpha}y(z). \label{W-scale}
\end{equation}

Next we determine the scaling forms of the drift and diffusion
coefficients $D^{(1)}(x,t)$ and $D^{(2)}(x,t)$. Following the same
way of determining the transformation between $W(x,t)$ and $y(z)$,
we can have $D^{(1)}(x,t)=(t^{\delta_2}/x^{\lambda_2})\rho_1(z)$.
The scaling behavior for $D^{(1)}(x,t)$ leads to
$d=b\delta_2-a\lambda_2$. For simplicity, we set $\lambda_2=0$.
This gives $\delta_2=\alpha-1$, where the relations between the
scaling exponents, namely, $b=a-d$ and $\alpha=a/b$, have been
used. A similar procedure is applied to determine $D^{(2)}(x,t)$
in terms of $\rho_2(z)$. To summarize, the scaling forms of the
coefficients are
\begin{align}\label{E3.3}
D^{(1)}(x,t)=t^{\alpha-1}\rho_1(z)\;\;\;,\;\;\;D^{(2)}(x,t)=t^{2\alpha-1}\rho_2(z)\;.
\end{align}

With eqs.~(\ref{E3.1}), (\ref{E3.2}) and (\ref{E3.3}), the FPE is
reduced to
\begin{align}\label{E3.4}
\rho_2(z)\,y''(z)+\Big[2\rho_2'(z)-\rho_1(z)+\alpha z\Big]\,y'(z)
+\Big[\rho_2''(z)-\rho_1'(z)+\alpha \Big]\,y(z)=0\;,
\end{align}
where the prime denotes the derivative with respect to $z$. Thus
it is seen that the solvability of the FPE, eq.~(\ref{E2.1}),
under the similarity method depends solely on that of
eq.~(\ref{E3.4}). An easy way to find exact solutions of
eq.~(\ref{E3.4}) is to relate it to either the hypergeometric
equation or the confluent hypergeometric equation. This requires
that $\rho_1(z)$ be a linear function of $z$, and $\rho_2(z)$ a
(linear) quadratic function of $z$ for the (confluent)
hypergeometric case.

However, it turns out that $\rho_1(z)$ and $\rho_2(z)$ need not be
so restricted in order to make eq.~(\ref{E3.4}) exactly solvable.
This is because eq.~(\ref{E3.4}) is exactly integrable.
Integrating it once, we get
\begin{equation}
\rho_2(z) y^\prime(z) + \left[\rho_2^\prime (z) - \rho_1 (z)+
\alpha z\right] y(z)=C, \label{y1}
\end{equation}
where $C$ is an integration constant.

To fix the constant $C$, we consider the boundary conditions of
the probability density $W(x,t)$ and the associated probability
current density $J(x,t)$.

\subsection{Relation between probability density and probability current density}

From the continuity equation
\begin{equation}
\frac{\partial}{\partial t}W(x,t)=-\frac{\partial}{\partial
x}\,J(x,t)\;,\label{E4.5.2}
\end{equation}
we have
\begin{equation}
J(x,t)=D^{(1)}(x,t)\,W(x,t)-\frac{\partial}{\partial
x}\,\Big[D^{(2)}(x,t)\,W(x,t)\Big]\;.\label{E4.5.3}
\end{equation}
Using eqs.~(\ref{E3.1}), (\ref{W-scale}) and (\ref{E3.3}), we get
\begin{equation}
J(x,t)=t^{-1} \left[\left(\rho_1(z)-\rho_2^\prime(z)\right)y(z)
-\rho_2 (z) y^\prime(z)\right]. \label{J-1}
\end{equation}
From eq.~(\ref{y1}), we can reduce the above equation to
\begin{equation}\label{E4.5.4a}
J(x,t)=\frac{1}{t}\left[\alpha\,z\,y(z)-C\right]=\frac{1}{t}\left[\alpha\,x\,W(x,t)-C\right],
\end{equation}
where $W(x,t)=t^{-\alpha}\,y(z)$ and $z=x/t^{\alpha}$ have been
used in obtaining the second expression.

For the domains which we are interested in, i.e., the whole line
and the half lines, the natural boundary conditions are
\begin{equation}
J(x,t)|_{\rm boundary}=0,~~~zy(z)|_{\rm boundary}=xW(x,t)|_{\rm
boundary}=0. \label{Conditions}
\end{equation}
These conditions imply that $C=0$, and that $J(x,t)$ is
proportional to $W(x,t)$ and $x$,
\begin{equation}\label{E4.5.4}
J(x,t)=\frac{1}{t}\alpha\,z\,y(z)=\frac{\alpha x}{t}\,W(x,t).
\end{equation}

Eq.~(\ref{E4.5.4}) can also be obtained by scaling consideration
as follows. Under the scale transformation, the scaling behavior
of $J(x,t)$ takes the form
$\bar{J}(\bar{x},\bar{t})=\varepsilon^{h}\,J(x,t)$, where $h=-b$
is determined from expression (\ref{E4.5.3}) with the help of
(\ref{E2.2}), (\ref{E2.3}), the relation $b=a-d=2a-e$ and $c=-a$.
The general scaling form of the probability current is
$J(x,t)=(t^{\delta_3}/x^{\lambda_3})\,\Sigma(z)$. Without loss of
generality, we choose the set of parameters
$(\lambda_3,\delta_3)=(0,-1)$.  This gives
\begin{align}\label{E4.5.6}
J(x,t)=t^{-1}\,\Sigma(z)\;.
\end{align}
Inserting the similarity solution $W(x,t)=t^{-\alpha}\,y(z)$ and
(\ref{E4.5.6}) into eq.~(\ref{E4.5.2}), one finds $\alpha
y(z)+\alpha z y'(z)=\Sigma'(z)$. This can be integrated to give
$\Sigma(z)=\int\,[\alpha y(z)+\alpha z
y'(z)]\,dz+\mbox{constant}=\alpha z y(z)+\mbox{constant}$.
Adopting the conditions (\ref{Conditions}) then gives
eq.~(\ref{E4.5.4}).

\subsection{Analytic expression of similarity solution $W(x,t)$}

With $C=0$, eq.~(\ref{y1}) is reduced to
\begin{equation}
\rho_2 (z) y^\prime(z) + \left[\rho_2^\prime (z)- \rho_1 (z)+
\alpha z\right] y(z)=0. \label{y2}
\end{equation}
This equation is easily solved to give
\begin{equation}
y(z)\propto \exp\left(\int^z dz\,
\frac{\rho_1(z)-\rho_2^\prime(z)-\alpha
z}{\rho_2(z)}\right),~~~\rho_2(z)\neq 0. \label{y-soln}
\end{equation}
Thus the probability density function $W(x,t)$ is given by
\begin{eqnarray}
W(x,t)&=&\mathcal{A} t^{-\alpha} \exp\left(\int^z
dz\,f(z)\right)_{z=\frac{x}{t^\alpha}},\nonumber\\
f(z)&\equiv& \frac{\rho_1(z)-\rho_2^\prime(z)-\alpha
z}{\rho_2(z)},~~~\rho_2(z)\neq 0,\label{W-soln}
\end{eqnarray}
where $\mathcal{A}$ is the normalization constant.  It is
interesting to see that the similarity solution of the FPE can be
given in such an analytic closed form.  Exact similarity solutions
of the FPE can be obtained as long as $\rho_1(z)$ and $\rho_2(z)$
are such that the function $f(z)$ in eq.~(\ref{W-soln}) is an
integrable function and the resulted $W(x,t)$ is normalizable.
Equivalently, for any integrable function $f(z)$ such that
$W(x,t)$ is normalizable, if one can find a function $\rho_2(z)$
($\rho_1(z)$ is then determined by $f(z)$ and $\rho_2(z)$), then
one obtains an exactly solvable FPE with similarity solution given
by (\ref{W-soln}).

Our discussions so far indicate a general way to construct exactly
solvable FPE with similarity solutions.  We shall not attempt to
exhaust all possibilities here.  Instead, we will only present a
few interesting cases.  We will mainly be concerned with cases
related to $f(z)$ having the forms (i) $f(z)=Az+B$ and (ii)
$f(z)=A+B/z$, where $A$ and $B$ are real constants. The
corresponding $W(x,t)$ are
\begin{eqnarray}
W(x,t) &\propto&  t^{-\alpha}e^{\frac{A}{2}z^2+Bz},\nonumber\\
W(x,t) &\propto&  t^{-\alpha} z^B\,e^{Az},
\end{eqnarray}
respectively.  Possible choices of $\rho_1 (z)$ and $\rho_2 (z)$
that give $f(z)$ in form (i) and (ii) are, respectively, (i)
$\rho_1(z)=\mu_1 z+\mu_2$, $\rho_2(z)=\mu_4$, and (ii)
$\rho_1(z)=\mu_1z+\mu_2$, $\rho_2(z)=\mu_3 z$.  We will discuss
these two cases in Sect.~IV and V.

\section{FPE with $\rho_1(z)=\mu_1 z+\mu_2$ and $\rho_2(z)=\mu_4$}

Let us take $\rho_1(z)$ and $\rho_2(z)$ as
\begin{align}\label{E4.2}
\rho_1(z)=\mu_1 z+\mu_2\;\;\;,\;\;\;\rho_2(z)=\mu_4\;,
\end{align}
where $\mu_1$, $\mu_2$ and $\mu_4$ are real constants. This choice
of $\rho_1 (z)$ and $\rho_2 (z)$ generate the following drift and
diffusion coefficients:
\begin{align}\label{E4.3}
D^{(1)}(x,t)=\mu_1\,\frac{x}{t}+\mu_2\,t^{\alpha-1}\;\;\;,\;\;\;
D^{(2)}(x,t)=\mu_4\,t^{2\alpha-1}\;.
\end{align}
From eq.~(\ref{W-soln}), the function $y(z)$ is
\begin{eqnarray}
y(z) &\propto& \left\{
\begin{array}{ll}
\mbox{exp}\Big\{\frac{1}{\mu_4}\left[\left(\mu_1-\alpha\right)\frac{z^2}{2}+\mu_2 z\right]\Big\}, &\mu_1\neq \alpha\\
& \\
\mbox{exp}\Big\{\frac{\mu_2}{\mu_4}z\Big\},  &\mu_1=\alpha
\end{array}
\right.
\label{case-I}
\end{eqnarray}
We shall discuss these two cases separately.

\subsection{$\mu_1 \neq \alpha$}

For this case, the normalized solution, from eq.~(\ref{W-soln}),
are
\begin{align}\label{E4.5}
W(x,t)=\sqrt{\frac{\alpha-\mu_1}{2\pi\mu_4 t^{2\alpha}}}\,
\mbox{exp}\Big\{-\frac{\alpha-\mu_1}{2\mu_4 t^{2\alpha}}
\Big(x-\frac{\mu_2 t^{\alpha}}{\alpha-\mu_1}\Big)^2\Big\}\;,
\end{align}
where either $(\mu_4>0$, $\mu_1<\alpha)$ or $(\mu_4<0$, $\mu_1>\alpha)$ must
be satisfied. The well-known diffusion equation is in this class
with $\alpha=1/2$, $\mu_1=\mu_2=0$ and $\mu_4>0$.

The solution~(\ref{E4.5}) is a Gaussian (or normal) distribution
in $x$. The full width at half maximum~(FWHM) of the solution is
related to the parameters $\alpha$, $\mu_1$, $\mu_4$ and time $t$
as $C_1\,\sqrt{(\mu_4 t^{2\alpha})/(\alpha-\mu_1)}$, where
$C_1=2\sqrt{2\ln 2}$. Hence, from eq.~(\ref{E4.3}), one can see
that the FWHM is affected by the coefficient $\mu_1$ of the term
$x/t$ of the drift coefficient and the coefficient $\mu_4$ of the
term $t^{2\alpha-1}$ of the diffusion coefficient respectively.
Also, it can be seen that the location of the peak of the
solution, $x=(\mu_2 t^{\alpha})/(\alpha-\mu_1)$, depends on the
parameters $\alpha$, $\mu_1$, $\mu_2$ and time $t$. From
eq.~(\ref{E4.3}), it is seen that the location of the peak of the
probability density can only be influenced by the drift
coefficient. The parameter $\mu_2$ in the drift coefficient plays
an important role in the determination of the location of the
peak.  For instance, if $\mu_2=0$, then the peak stays at the
origin and will not change with time.  The value of the peak is
$\sqrt{(\alpha-\mu_1)/(2\pi\mu_4 t^{2\alpha})}$, and is dependent
on the parameters $\alpha$, $\mu_1$, $\mu_4$ and time $t$. Thus it
is affected by both the drift and diffusion coefficients.

One notes that at fixed time $t$, $W(x,t)$ with $\mu_2$ and
$-\mu_2$ are the mirror images to each other with respect to the
$y$-axis as the rest of the parameters are kept fixed.
Furthermore, when the parameters $(2\alpha-\mu_1, -\mu_4)$ take
the place of $(\mu_1, \mu_4)$ with the rest unchanged, the
corresponding two $W(x,t)$'s are the mirror images to each other.
One then finds that $W(x,t)$ is invariant as the parameters
$(\mu_1, \mu_2, \mu_4)$ are replaced by $(2\alpha-\mu_1, -\mu_2,
-\mu_4)$.

As shown previously, the FWHM and the peak value are both related
to $(\mu_4t^{2\alpha})/(\alpha-\mu_1)$. The parameter $\alpha$ can
be either positive or negative. When $\alpha>0$, the FWHM of the
solution (\ref{E4.5}) is getting broader, the peak is turning
smaller and is moving away from the origin with time. On the other
hand, if $\alpha<0$, then the FWHM of the solution will shrink,
the peak value will become higher and move toward the origin as
time elapses.  The situation where the FWHM is neither expanding
nor contracting is impossible because $\alpha\neq 0$.

\begin{figure}[ht] \centering
\includegraphics*[width=3.3cm,height=3.3cm]{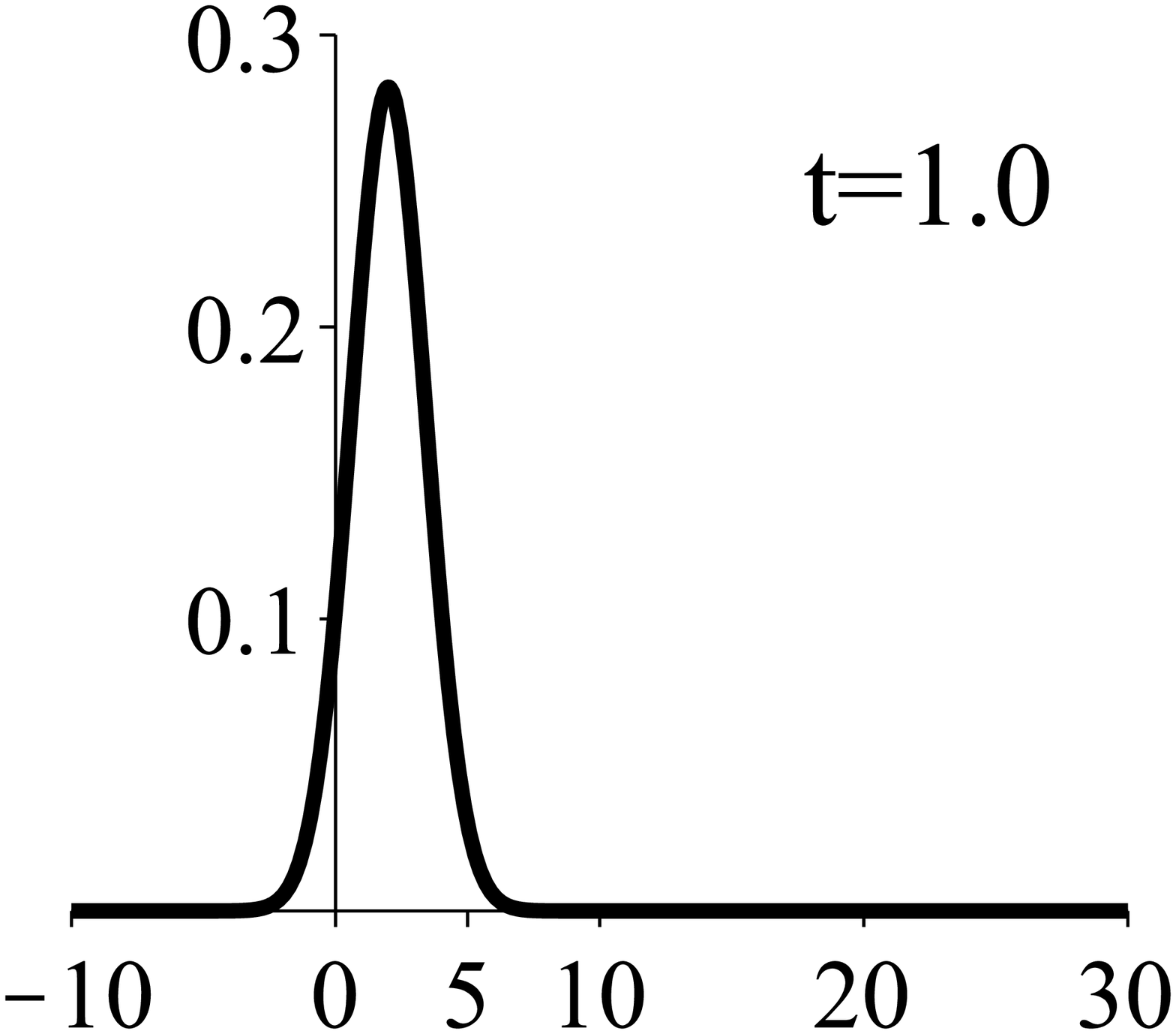}\hspace{3cm}
\includegraphics*[width=3.3cm,height=3.3cm]{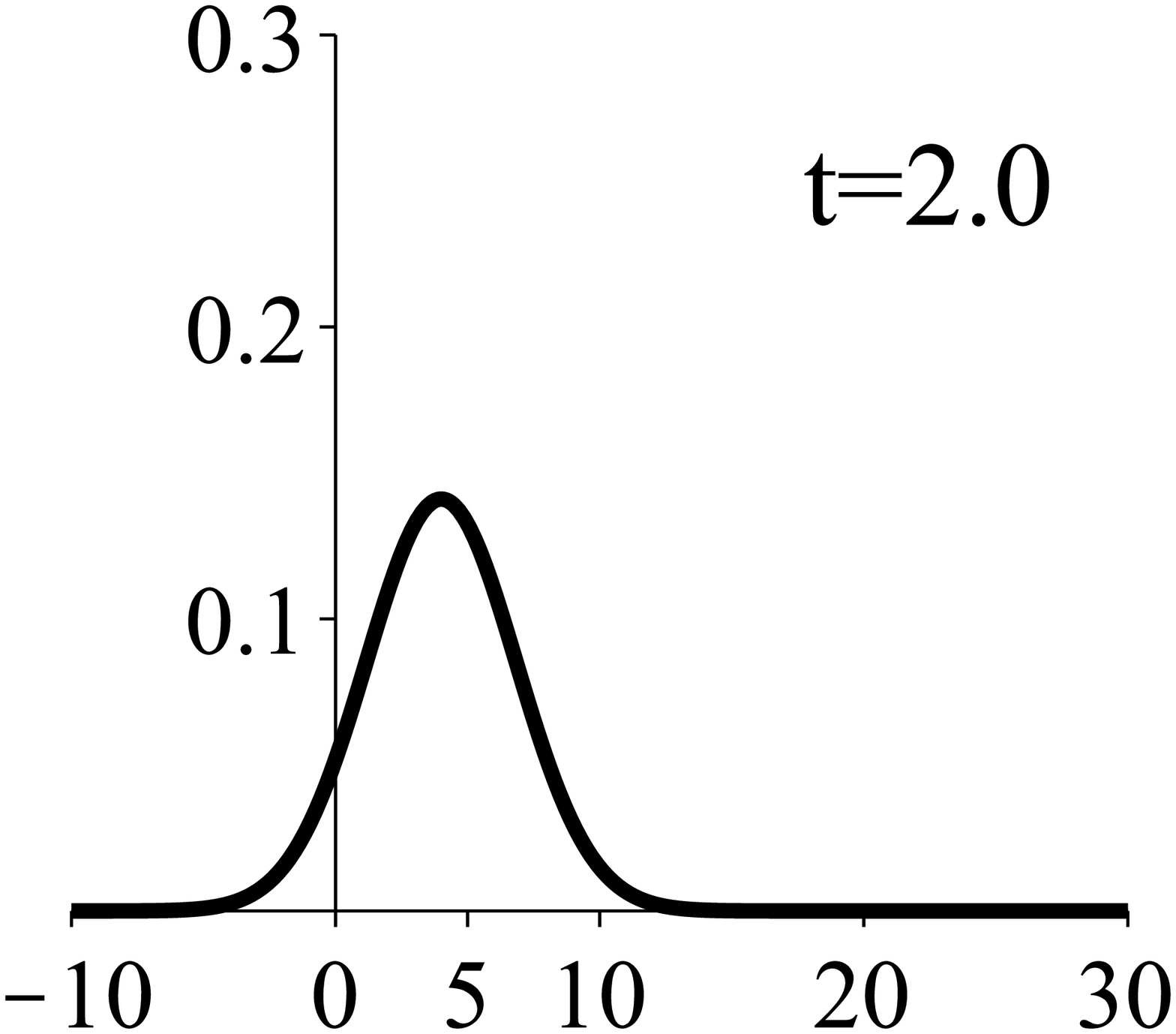}\\
\includegraphics*[width=3.3cm,height=3.3cm]{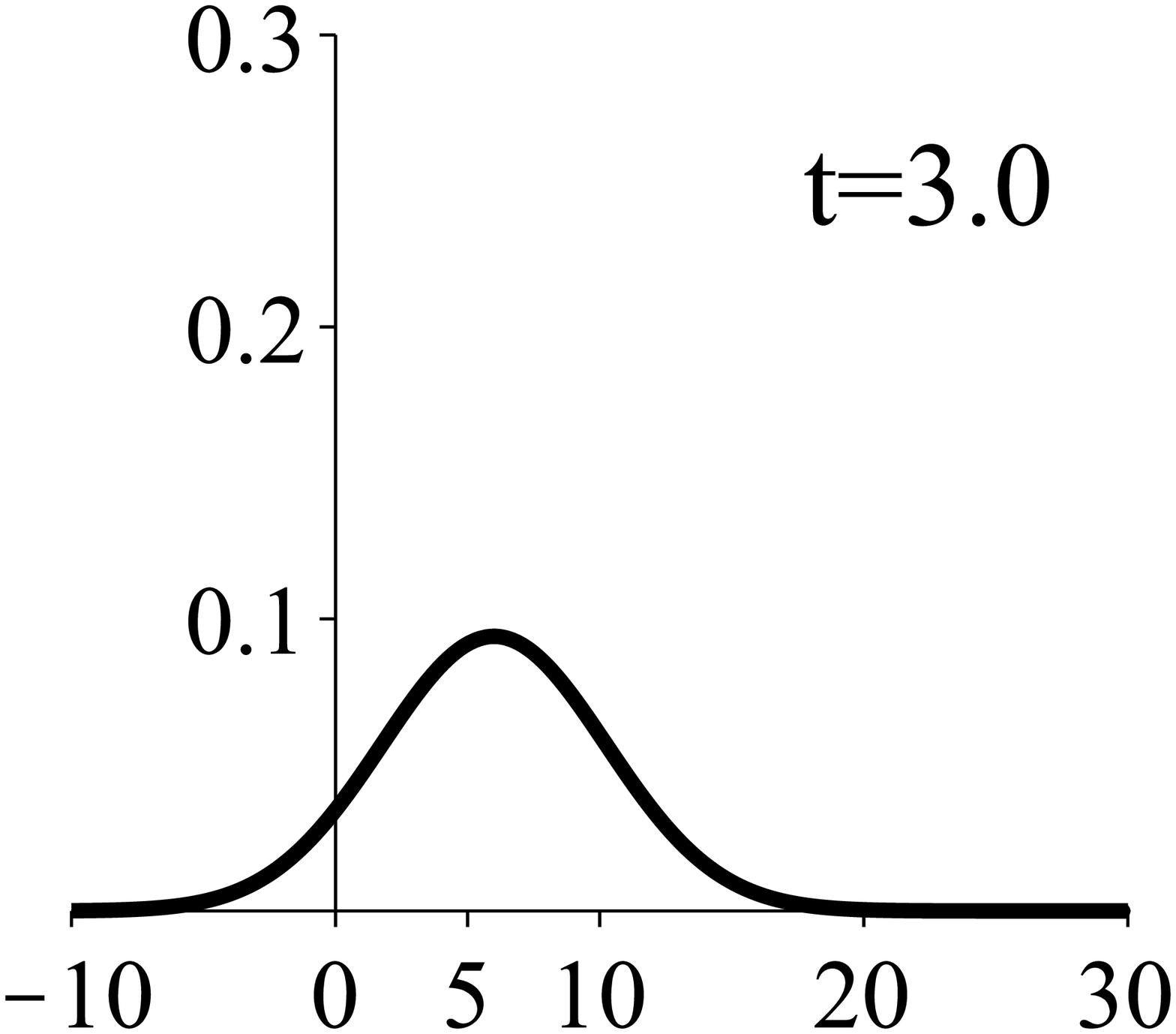}\hspace{3cm}
\includegraphics*[width=3.3cm,height=3.3cm]{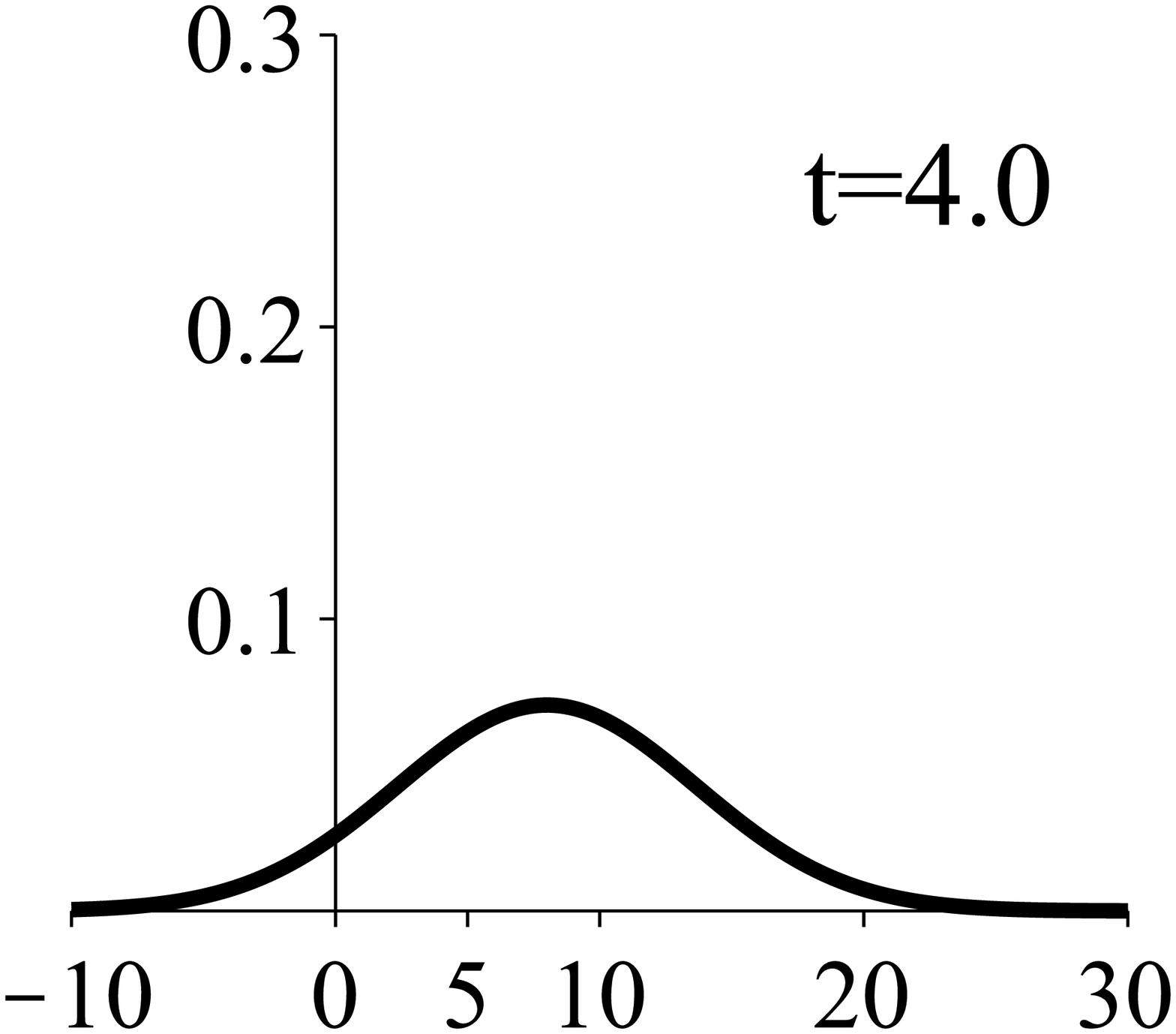}
\caption{Plot of $W(x,t)$ versus $x$ for solution~(\ref{E4.5})
with $\alpha=1$, $\mu_1=1/2$, $\mu_2=1$, $\mu_4=1$ and time
$t=1.0$, $2.0$, $3.0$, $4.0$.} \label{fig:EX1}
\end{figure}

FIG.~\ref{fig:EX1} shows the evolutions of the probability density
in solution~(\ref{E4.5}) for a set of the parameters, where
$\mu_1\neq\alpha$.

For the set of parameters taken in FIG.~\ref{fig:EX1}, the drift
and diffusion coefficient are $D^{(1)}(x,t)=x/(2t)+1$ and
$D^{(2)}(x,t)=t$, respectively. From the figures above and the
analysis in Sect. IV. A, one sees that the peak of the probability
distribution $W(x,t)$ moves to the right as time increases. This
is due to the presence of the drift force. Furthermore, owing to
the presence of $\mu_1$ and $\mu_4$ in the drift and diffusion
coefficient, the FWHM of $W(x,t)$ is getting wider, and the peak
value is getting smaller.

FIG.~\ref{fig:EX1} shows the evolution of the solution of the FPE
with a set of time-dependent drift and diffusion coefficient. In
subsections C and D, we shall study the difference in behavior of
the solutions as one of the coefficients changes from being
time-independent to being time-dependent, while the other
coefficient is being kept fixed.

\subsection{$\mu_1=\alpha$}

In this case, the normalized solution is
\begin{align}\label{E4.5.1}
W(x,t)=\Big|\frac{\mu_2}{\mu_4\,t^{\alpha}}\Big|\,\mbox{exp}\Big\{\frac{\mu_2}{\mu_4\,t^{\alpha}}\,x\Big\}\;,
\end{align}
where it is valid in $x\geq 0$ for $(\mu_2/\mu_4)<0$; $x\leq 0$ for $(\mu_2/\mu_4)>0$.

Solution (\ref{E4.5.1}) is the time-dependent solution of the FPE
with the time-dependent drift coefficient $D^{(1)}(x,t)=\alpha
x/t+\mu_2 t^{\alpha-1}$ and diffusion coefficient
$D^{(2)}(x,t)=\mu_4 t^{2\alpha-1}$.  It possesses certain symmetry
properties.  For instance, when the ratio of $\mu_2/\mu_4$ is
taken to be the same for different sets of $(\mu_2, \mu_4)$, the
solution~(\ref{E4.5.1}) is invariant. Also, when $(\mu_2, \mu_4)$
are replaced with $(-\mu_2, \mu_4)$ or $(\mu_2, -\mu_4)$, the
corresponding $W(x,t)$'s are the mirror images to each other with
the other parameters fixed.

In the following, one example is given to illustrate
the solution~(\ref{E4.5.1}) and the probability
current $J(x,t)$ of solution~(\ref{E4.5.1}).

\begin{figure}[ht] \centering
\includegraphics*[width=3.3cm,height=3.3cm]{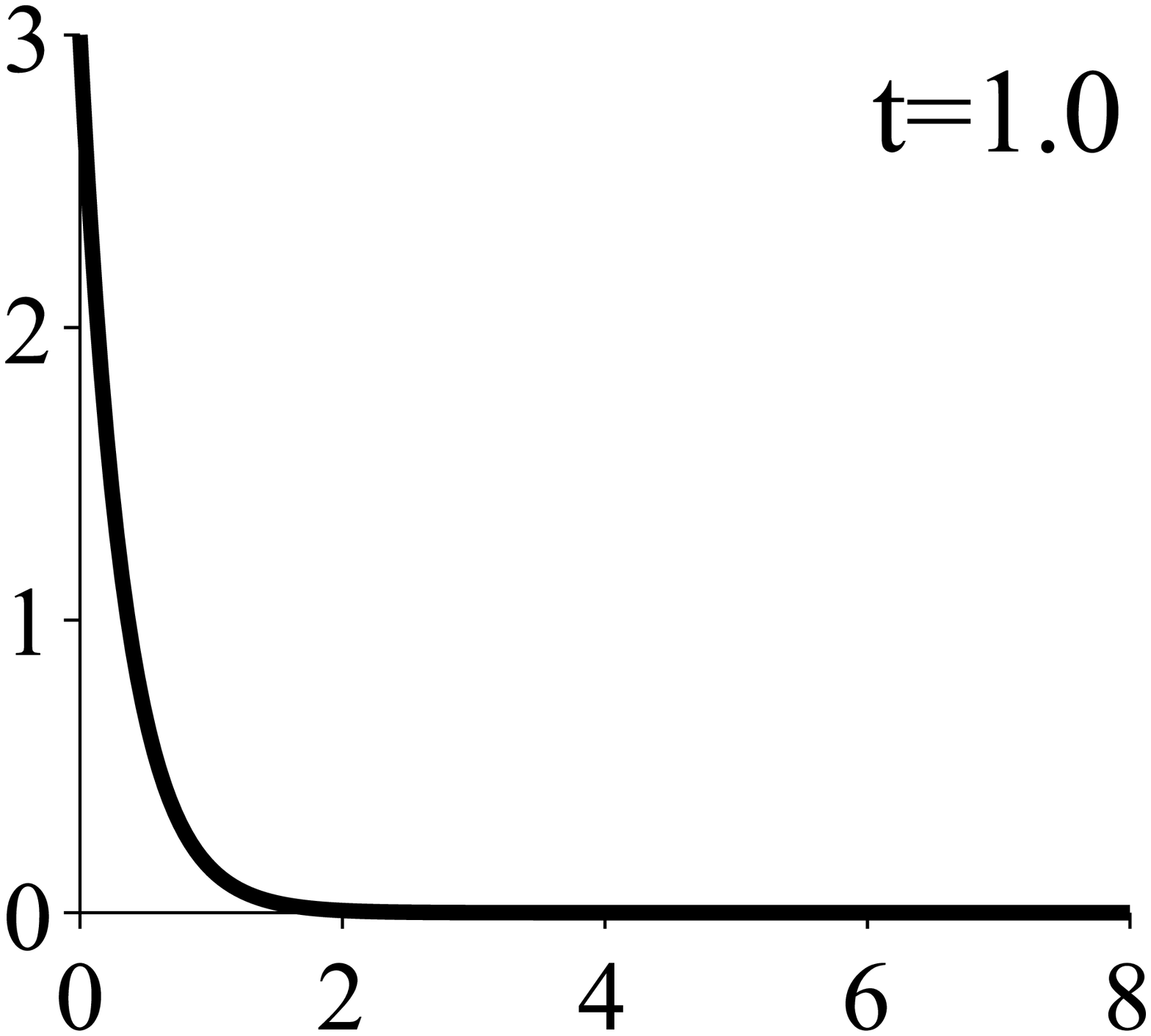}\hspace{3cm}
\includegraphics*[width=3.3cm,height=3.3cm]{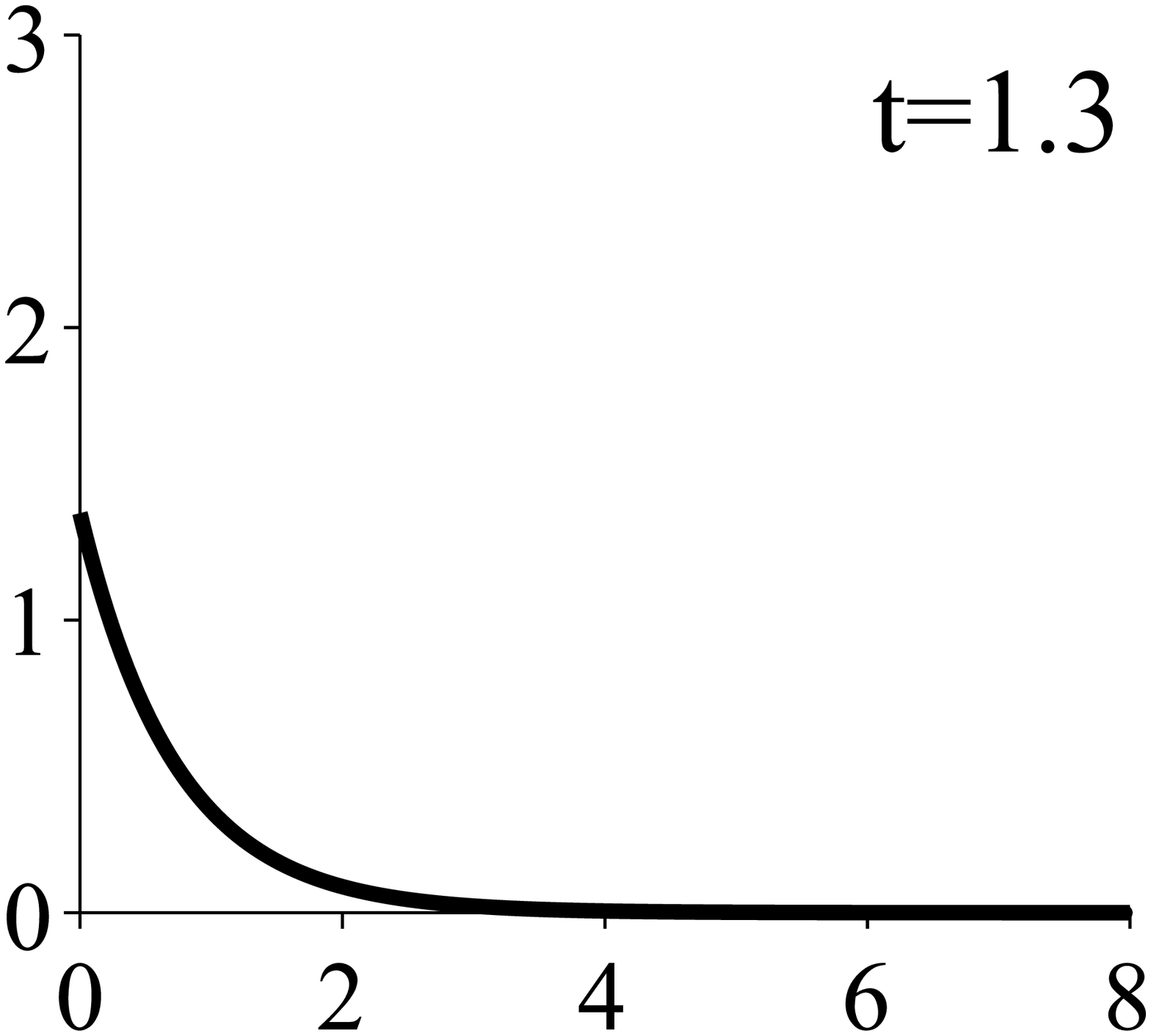}\\
\includegraphics*[width=3.3cm,height=3.3cm]{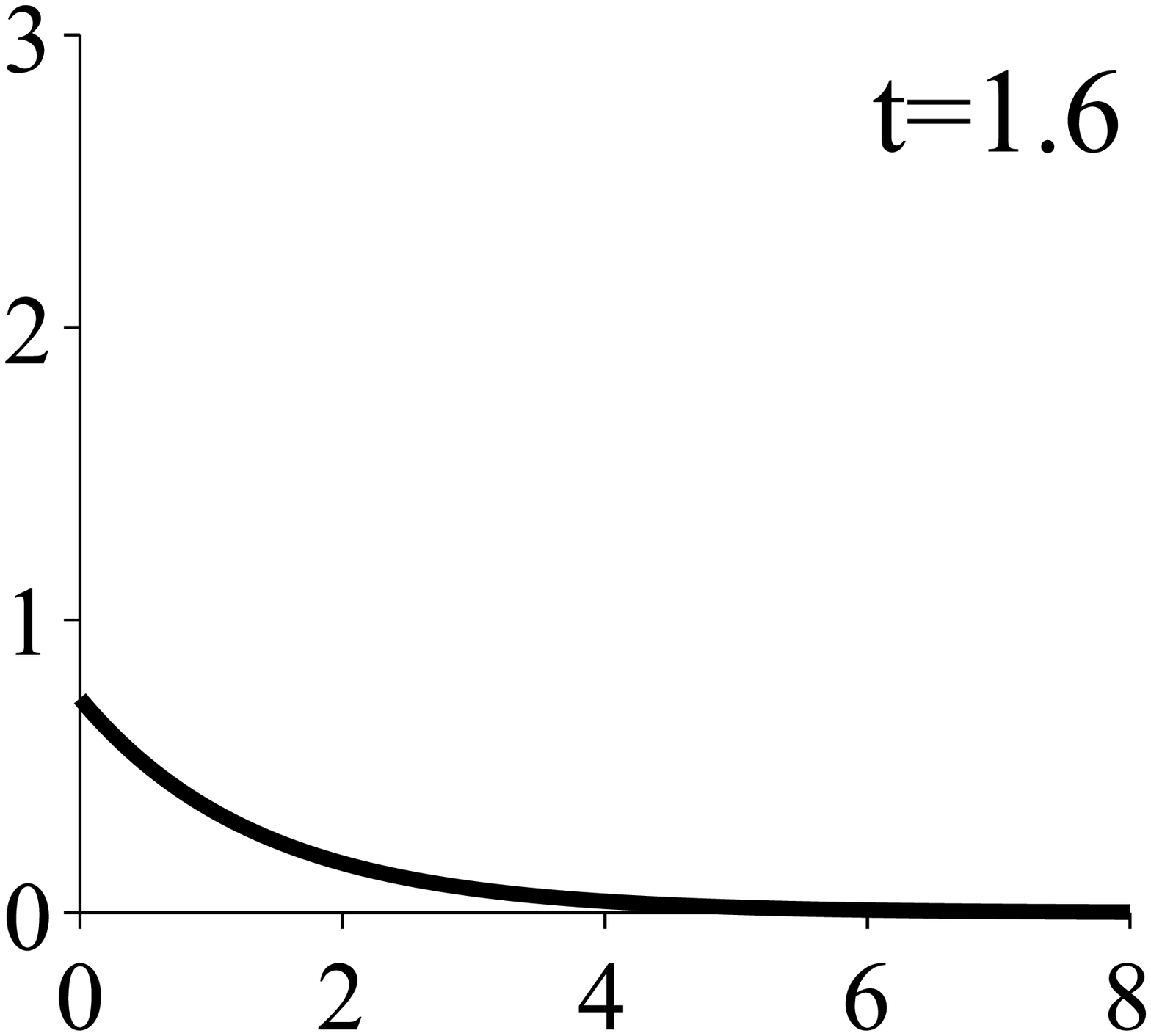}\hspace{3cm}
\includegraphics*[width=3.3cm,height=3.3cm]{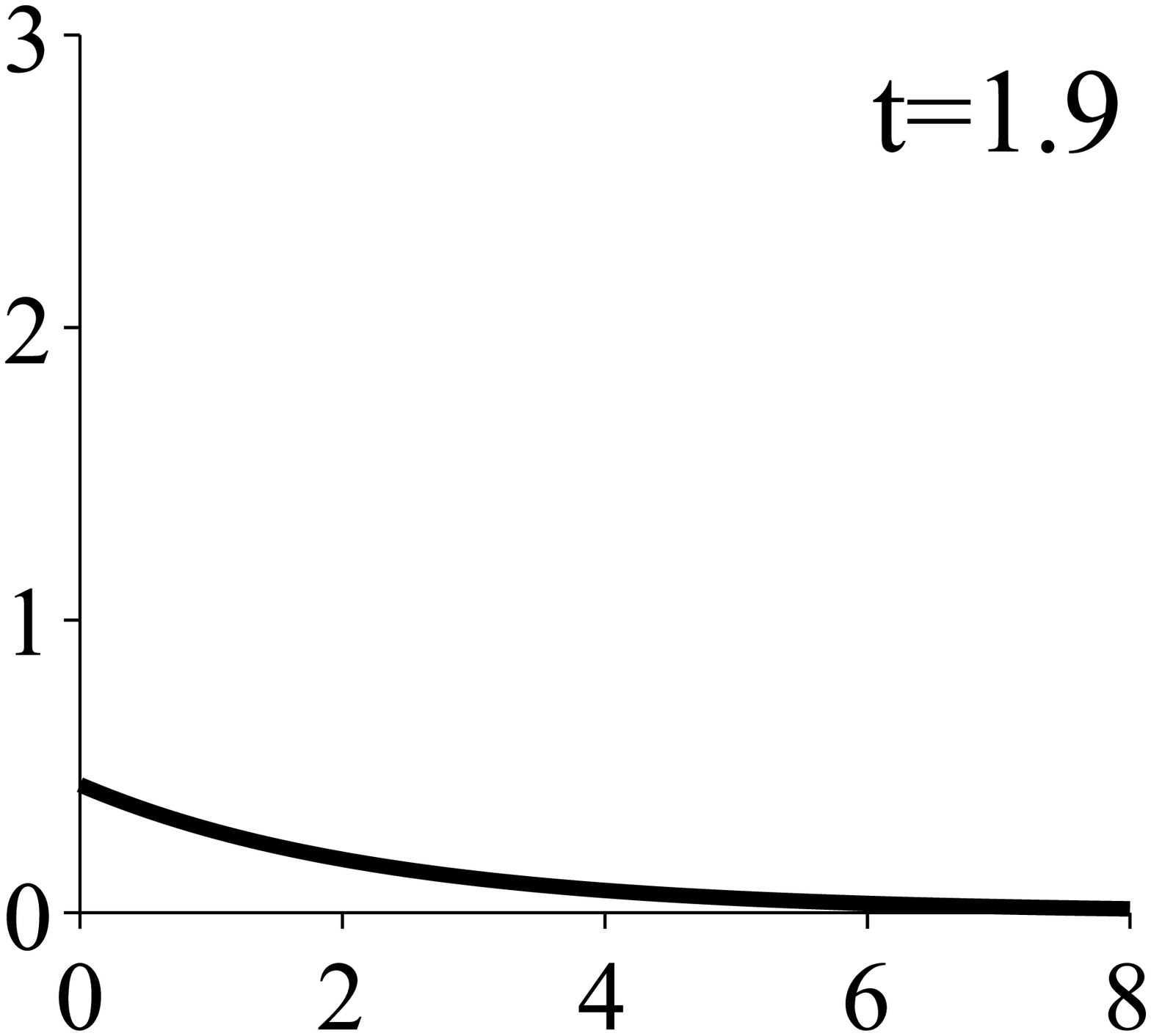}
\caption{Plot of $W(x,t)$ versus $x$ for solution~(\ref{E4.5.1})
with $\alpha=3$, $\mu_1=3$, $\mu_2=-6$, $\mu_4=2$ and time
$t=1.0$, $1.3$, $1.6$, $1.9$.} \label{fig:EX0}
\end{figure}

\begin{figure}[ht] \centering
\includegraphics*[width=3.3cm,height=3.3cm]{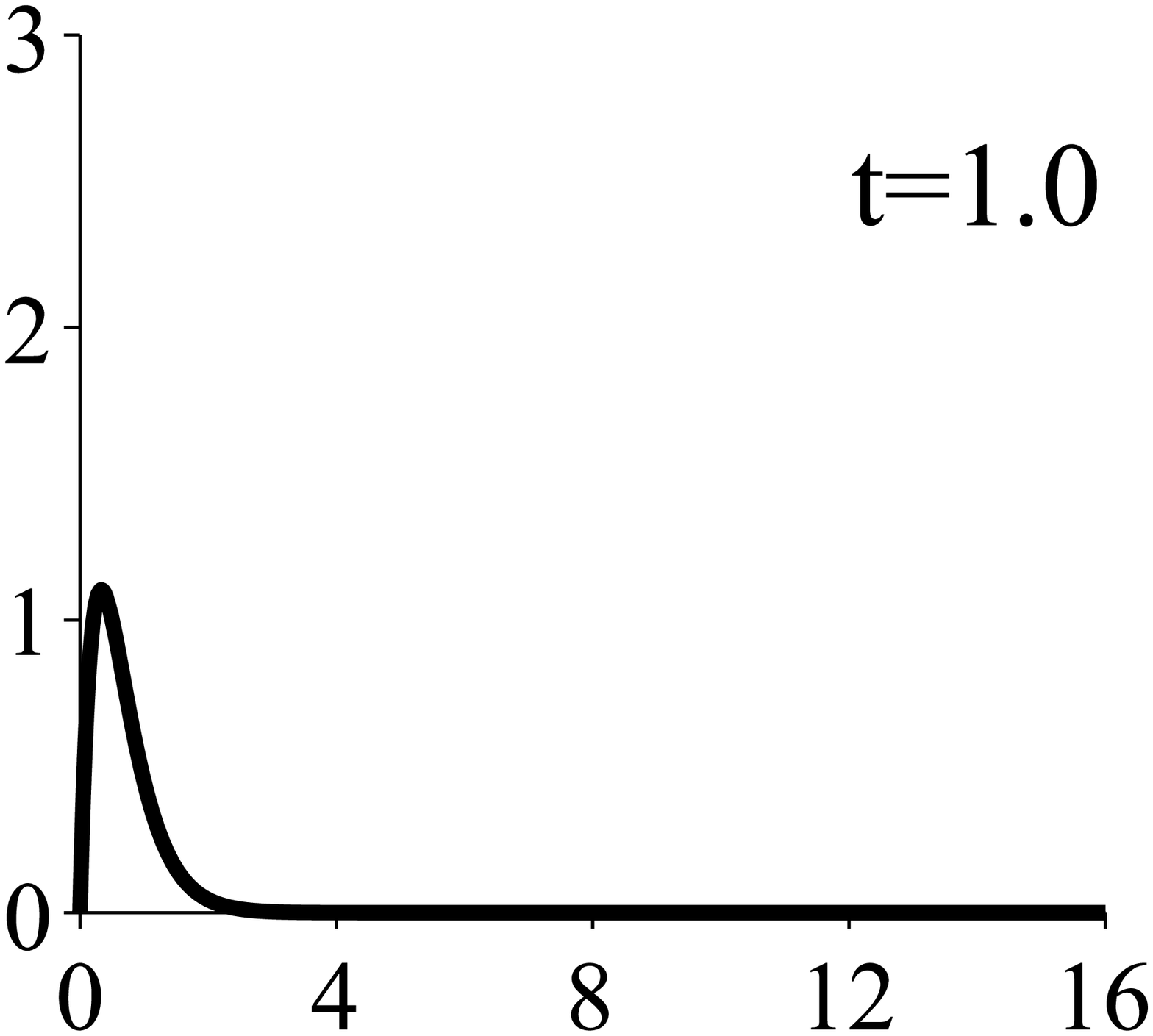}\hspace{3cm}
\includegraphics*[width=3.3cm,height=3.3cm]{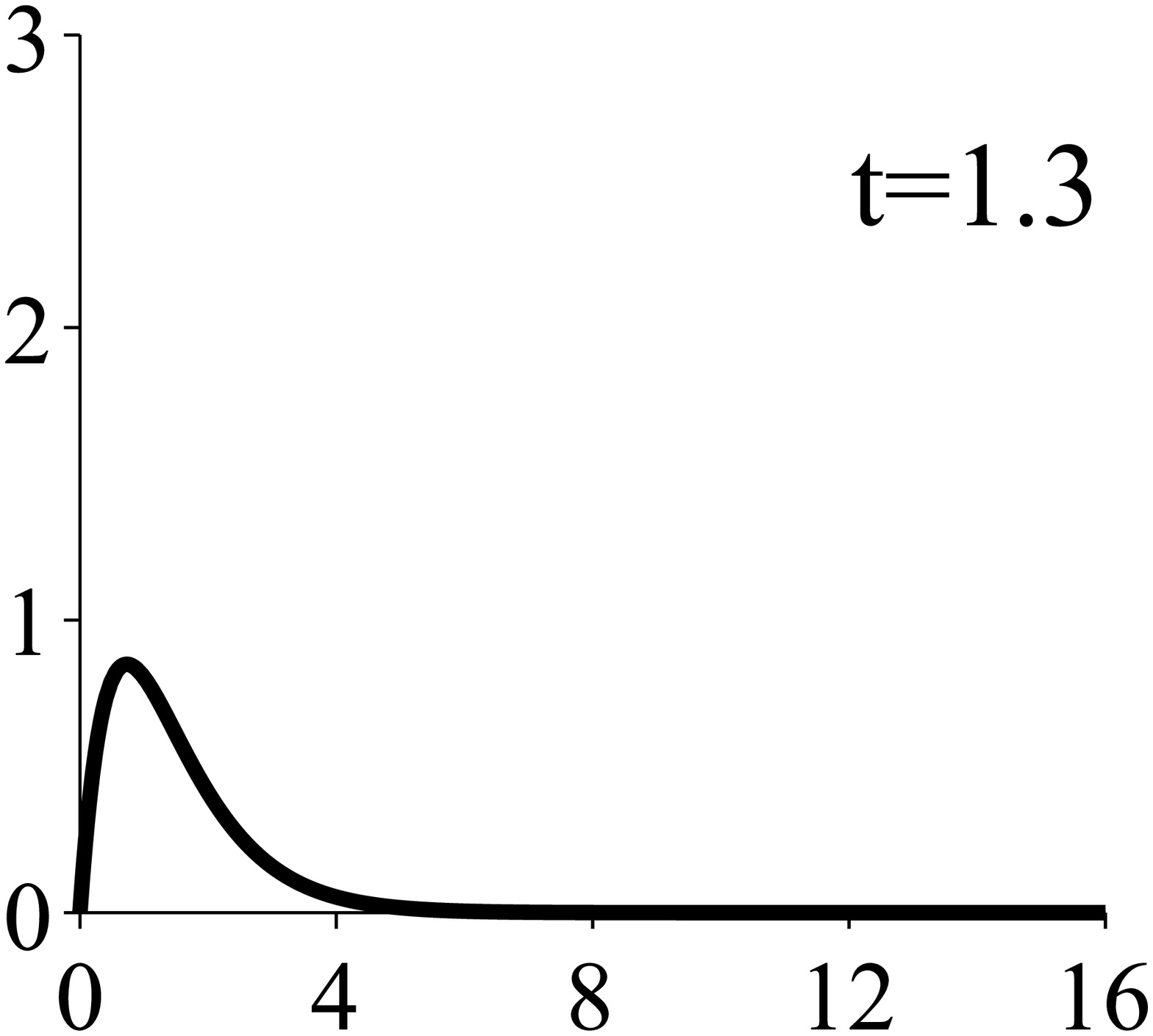}\\
\includegraphics*[width=3.3cm,height=3.3cm]{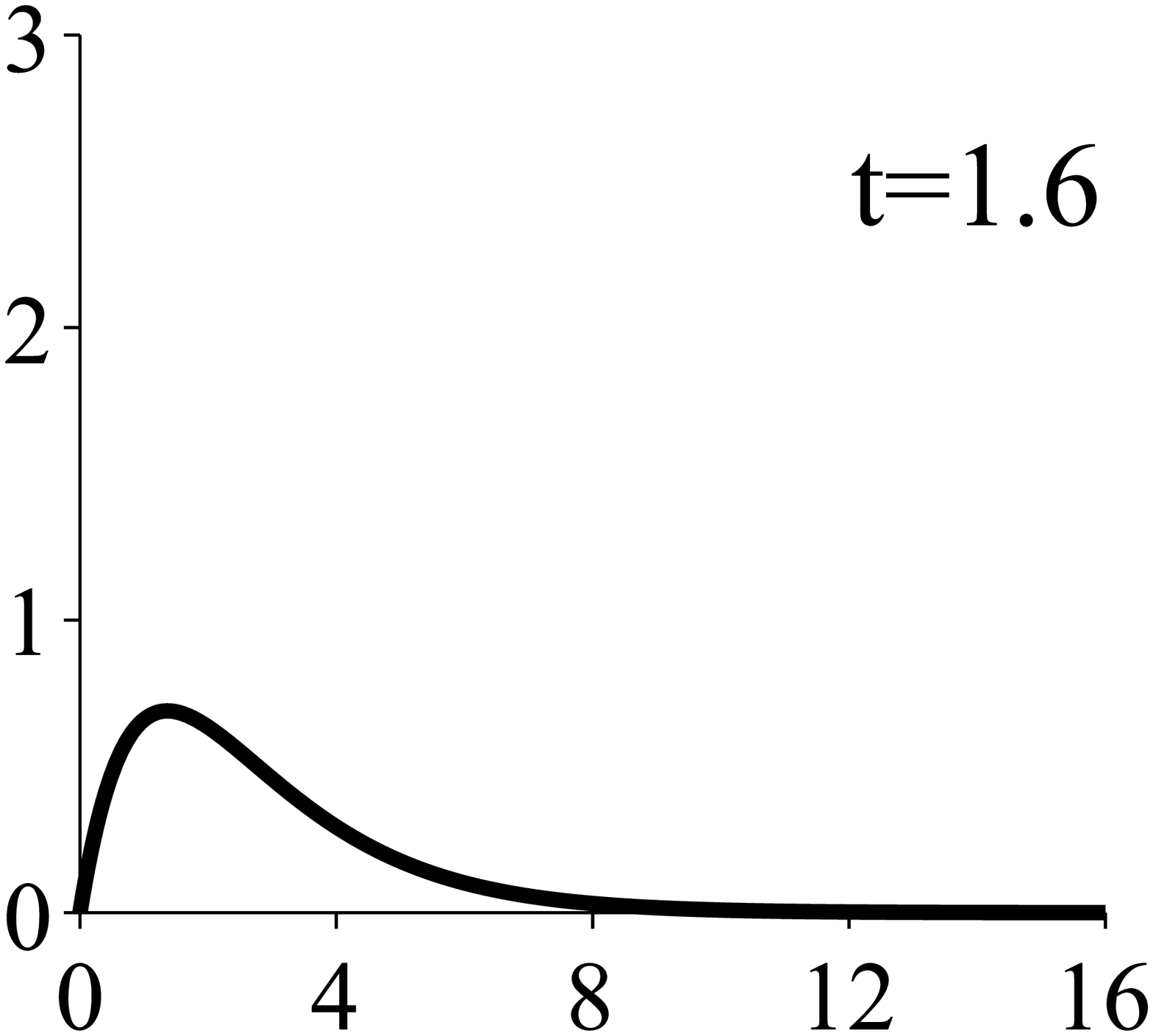}\hspace{3cm}
\includegraphics*[width=3.3cm,height=3.3cm]{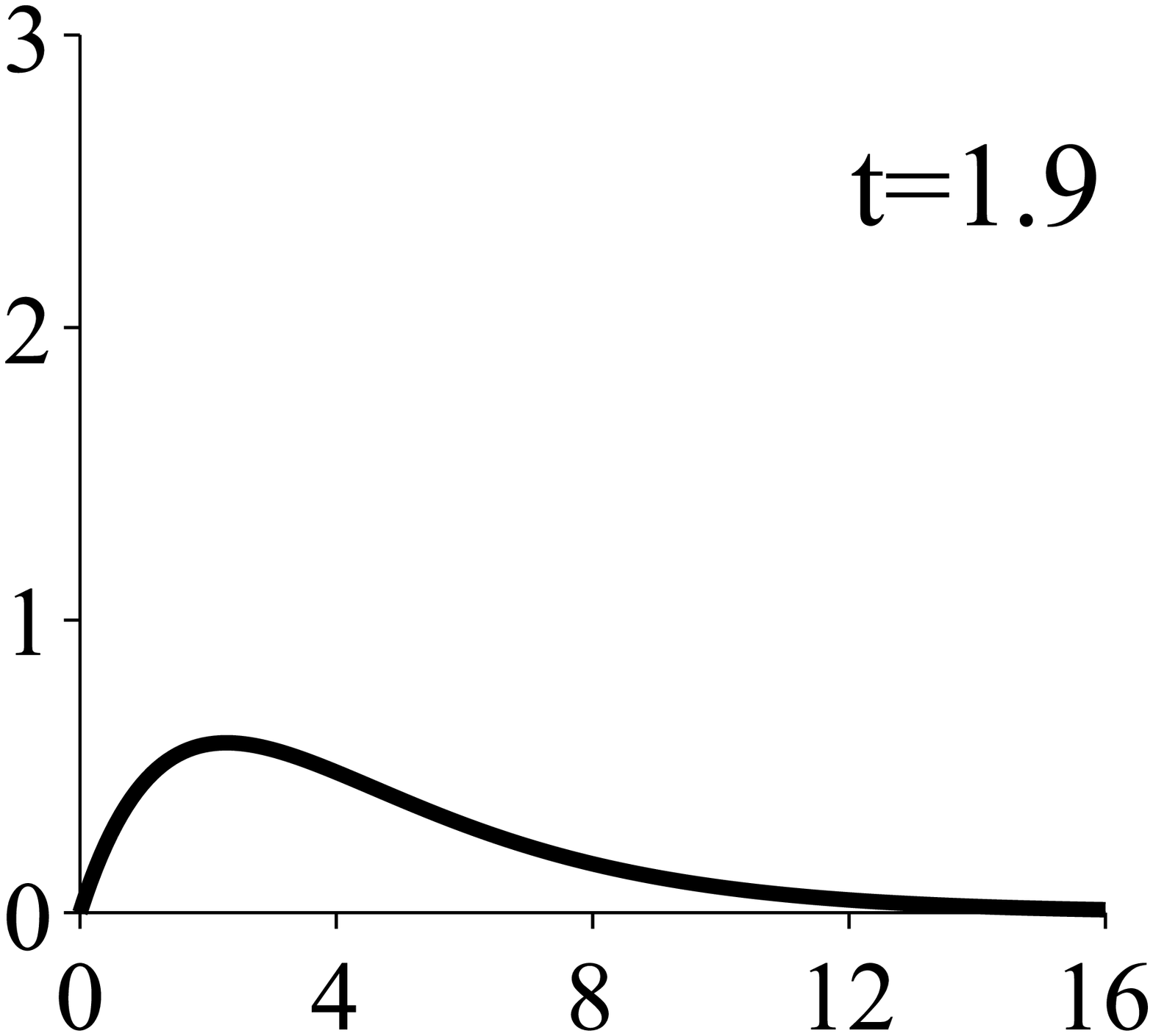}
\caption{Plot of $J(x,t)$ versus $x$ for solution~(\ref{E4.5.1})
with the same set of parameters as in FIG.~\ref{fig:EX0}.}
\label{fig:EX0-J}
\end{figure}

FIG.~\ref{fig:EX0} displays the evolutions of the probability
density in solution~(\ref{E4.5.1}) for a set of the parameters,
where $\mu_1=\alpha$. FIG.~\ref{fig:EX0-J} shows the probability
current $J(x,t)$ corresponding to the solution~(\ref{E4.5.1}) with
the same set of the parameters as in FIG.~\ref{fig:EX0}.

%\newpage
The time evolution of $W(x,t)$ in FIG.~\ref{fig:EX0} demonstrates
the decrease of the concentration near the origin and the spread
to the area away from the origin with time.

From (\ref{E4.5.4}) and (\ref{E4.5.1}), one can find that the
maximum probability current $J(x,t)$ occurs at $x=-\mu_4
t^{\alpha}/\mu_2$, which moves with time.

The plots in FIG.~\ref{fig:EX0} are the same for an infinite set
of $(\mu_2, \mu_4)$ with $\mu_2/\mu_4=-3$ with other parameters
remaining the same. It means that FPEs with drift coefficient
$D^{(1)}(x,t)=3x/t-3\mu_4t^2$ and diffusion coefficient
$D^{(2)}(x,t)=\mu_4t^5$  have the same solutions for arbitrary
$\mu_4\neq 0$.

\subsection{Examples with the same drift but different diffusion
coefficients}

Let us take $\mu_2=0$, and consider two different values of
$\alpha$, namely, $\alpha=1/2$ and $\alpha=1$.   The case with
$\alpha=1/2$ corresponds to $D^{(1)}(x,t)=\mu_1x/t$ and
$D^{(2)}(x,t)=\mu_4$, with the probability distribution
\begin{align}\label{E4.6}
W(x,t)=\sqrt{\frac{1-2\mu_1}{4\pi\mu_4 t}}\,\mbox{exp}
\Big\{-\Big(\frac{1-2\mu_1}{4\mu_4 t}\Big)\,x^2\Big\}\;.
\end{align}
The other case,  $\alpha=1$, leads to $D^{(1)}(x,t)=\mu_1x/t$,
$D^{(2)}(x,t)=\mu_4 t$ and
\begin{align}\label{E4.7}
W(x,t)=\sqrt{\frac{1-\mu_1}{2\pi\mu_4 t^2}}\,\mbox{exp}
\Big\{-\Big(\frac{1-\mu_1}{2\mu_4 t^2}\Big)\,x^2\Big\}\;.
\end{align}
These two examples have the same drift force, and differ only in
the diffusion coefficients: constant in the first case, and linear
in time in the second.  One can thus study how the time-dependent
diffusion modifies the behavior of the system with constant
diffusion.

As $\mu_2=0$, the peak of the probability distribution will not
move with time. Comparing eqs.~(\ref{E4.6}) and (\ref{E4.7}), one
sees that the FWHM is changed from $C_1\sqrt{(2\mu_4
t)/(1-2\mu_1)}$ to $C_1\sqrt{(\mu_4 t^2)/(1-\mu_1)}$, and the
value of the peak is changed from $\sqrt{(1-2\mu_1)/(4\pi\mu_4
t)}$ to $\sqrt{(1-\mu_1)/(2\pi\mu_4 t^2)}$.  The two distributions
coincide only at time $t_c=2(1-\mu_1)/(1-2\mu_1)$.

\begin{figure}[ht]
\centering
\includegraphics*[width=3.3cm,height=3.3cm]{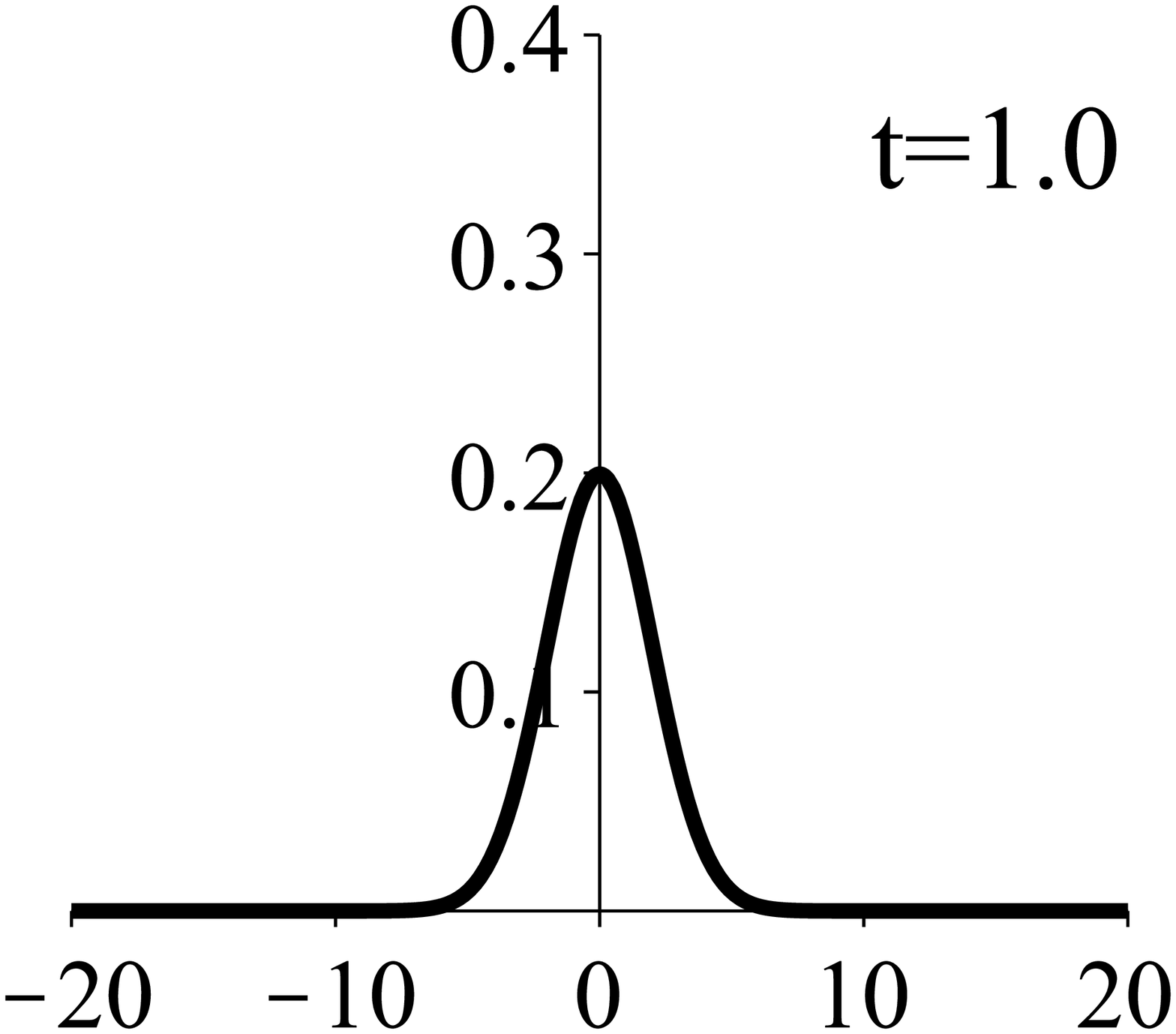}\hspace{3cm}
\includegraphics*[width=3.3cm,height=3.3cm]{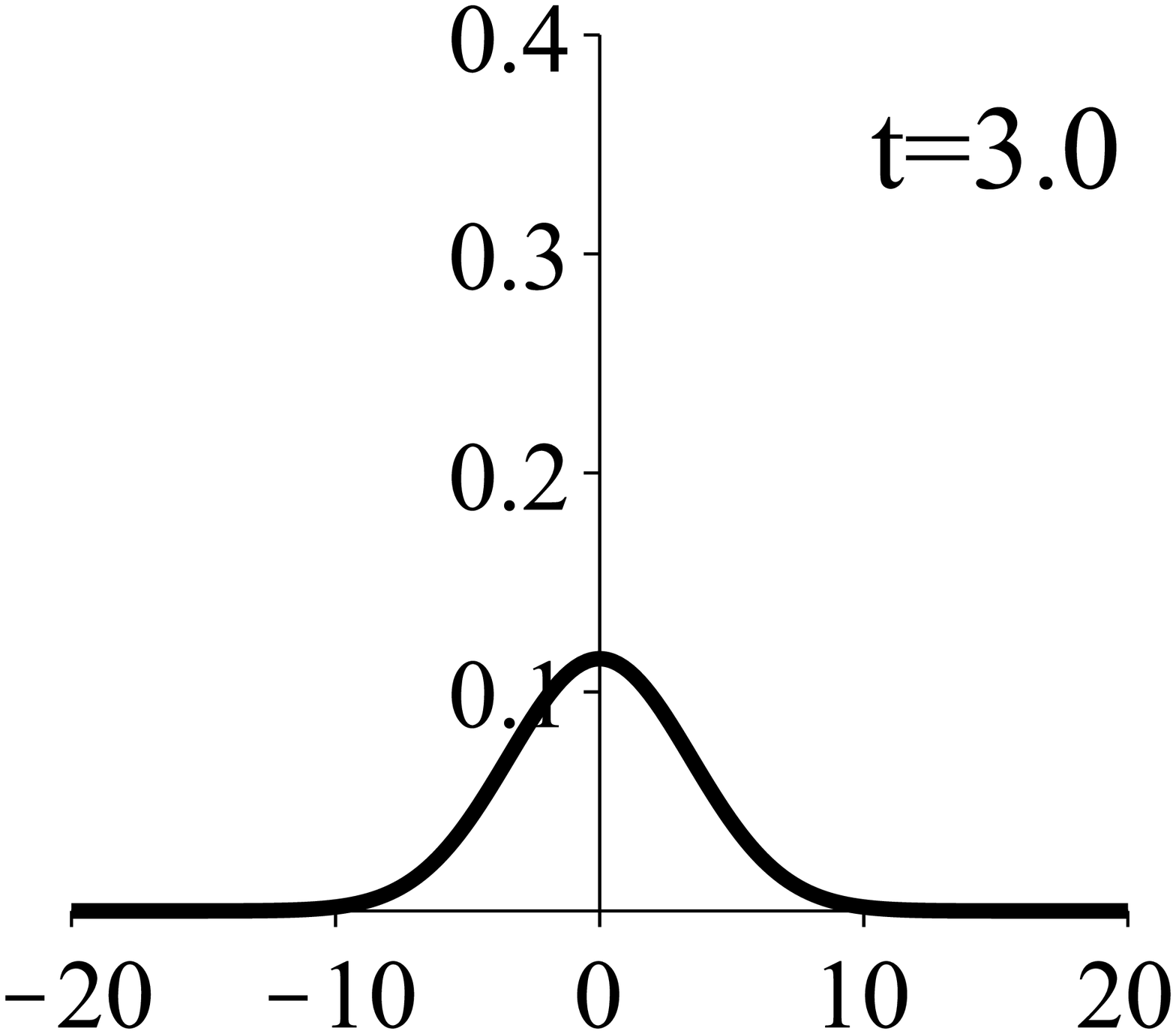}\\
\includegraphics*[width=3.3cm,height=3.3cm]{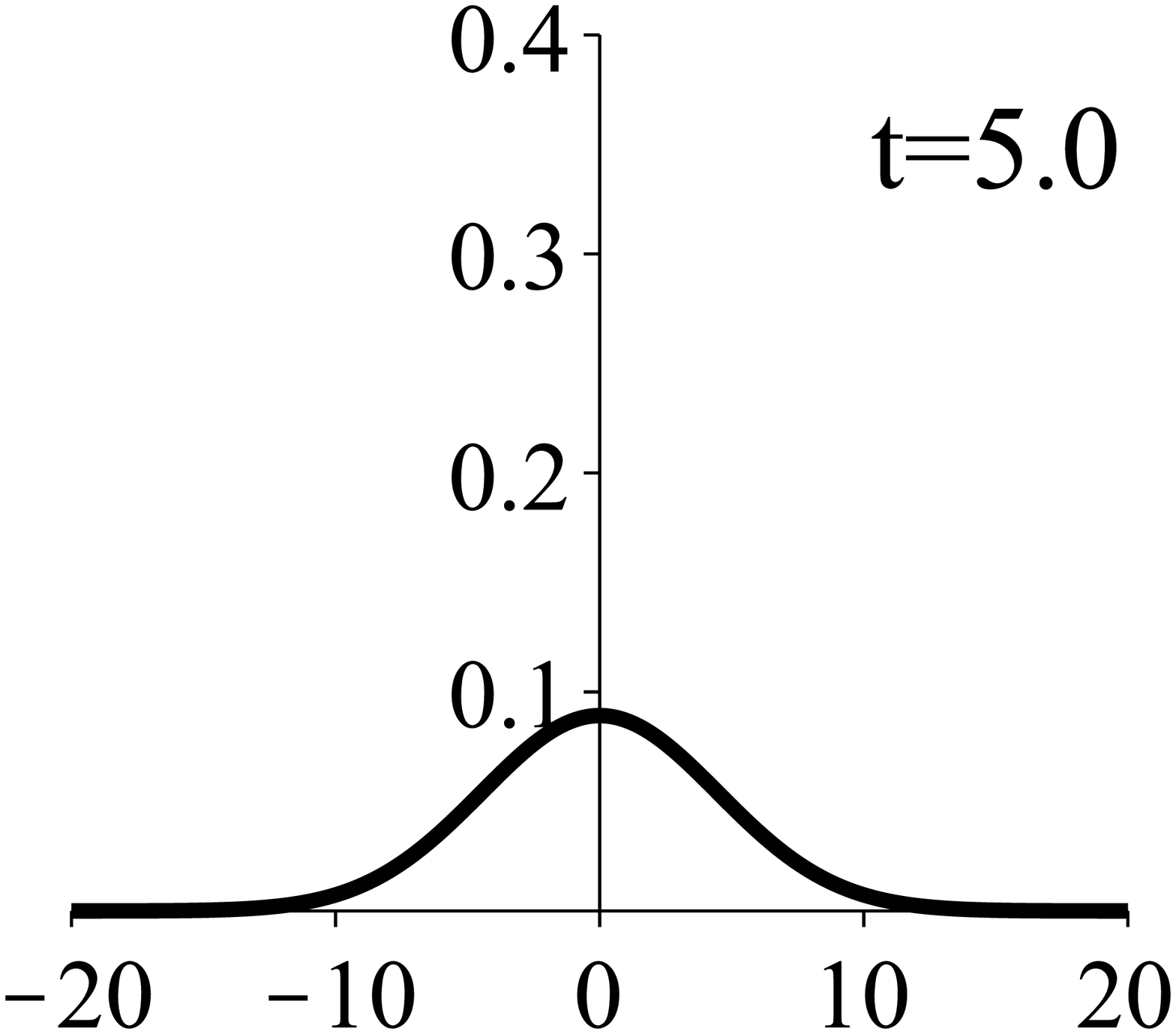}\hspace{3cm}
\includegraphics*[width=3.3cm,height=3.3cm]{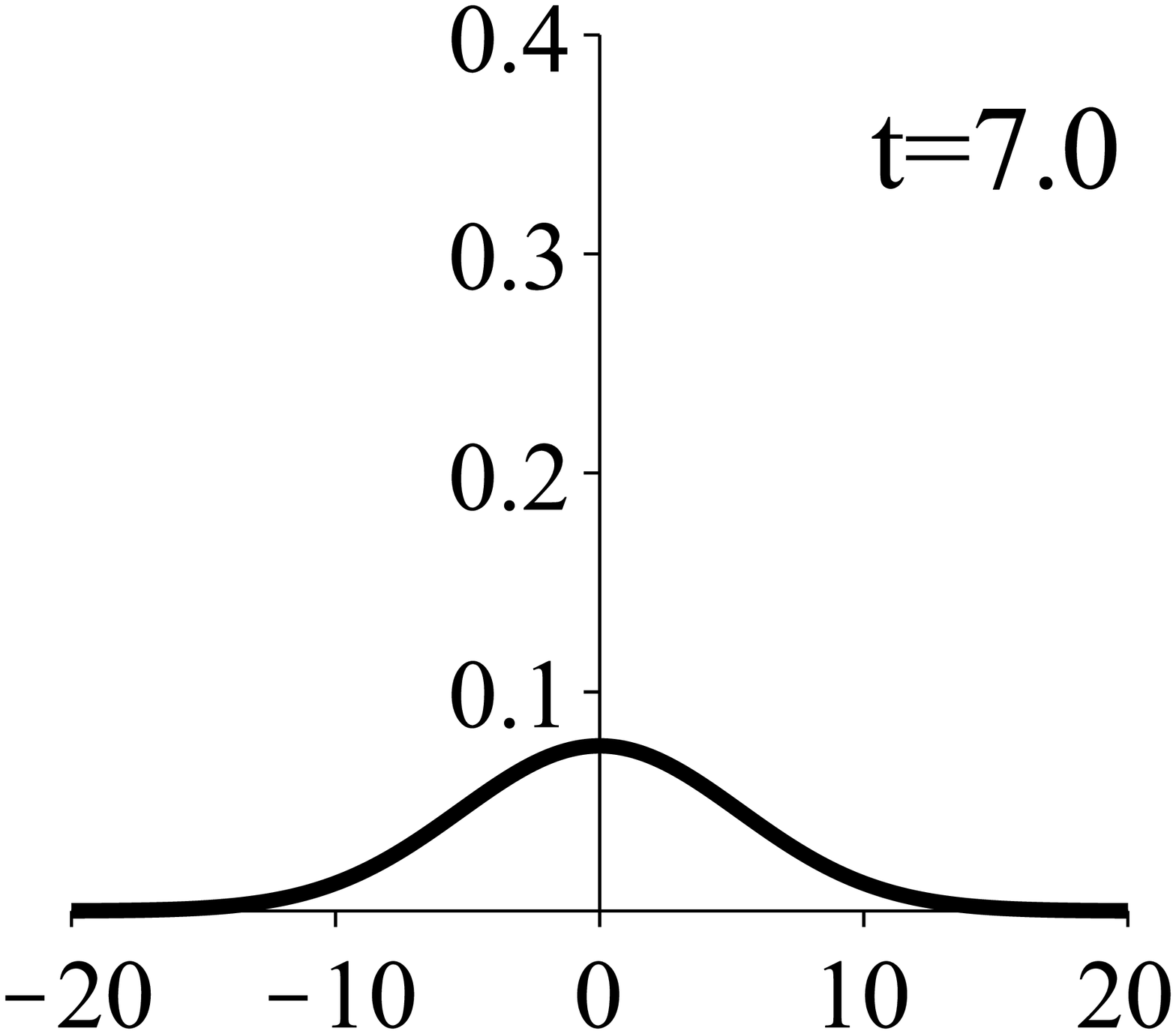}
\caption{Plot of $W(x,t)$ versus $x$ for solution~(\ref{E4.6})
with $\alpha=1/2$, $\mu_1=1/4$, $\mu_2=0$, $\mu_4=1$, and time
$t=1.0$, $3.0$, $5.0$, $7.0$.}\label{fig:EX2}
\end{figure}

\begin{figure}[ht]
\centering
\includegraphics*[width=3.3cm,height=3.3cm]{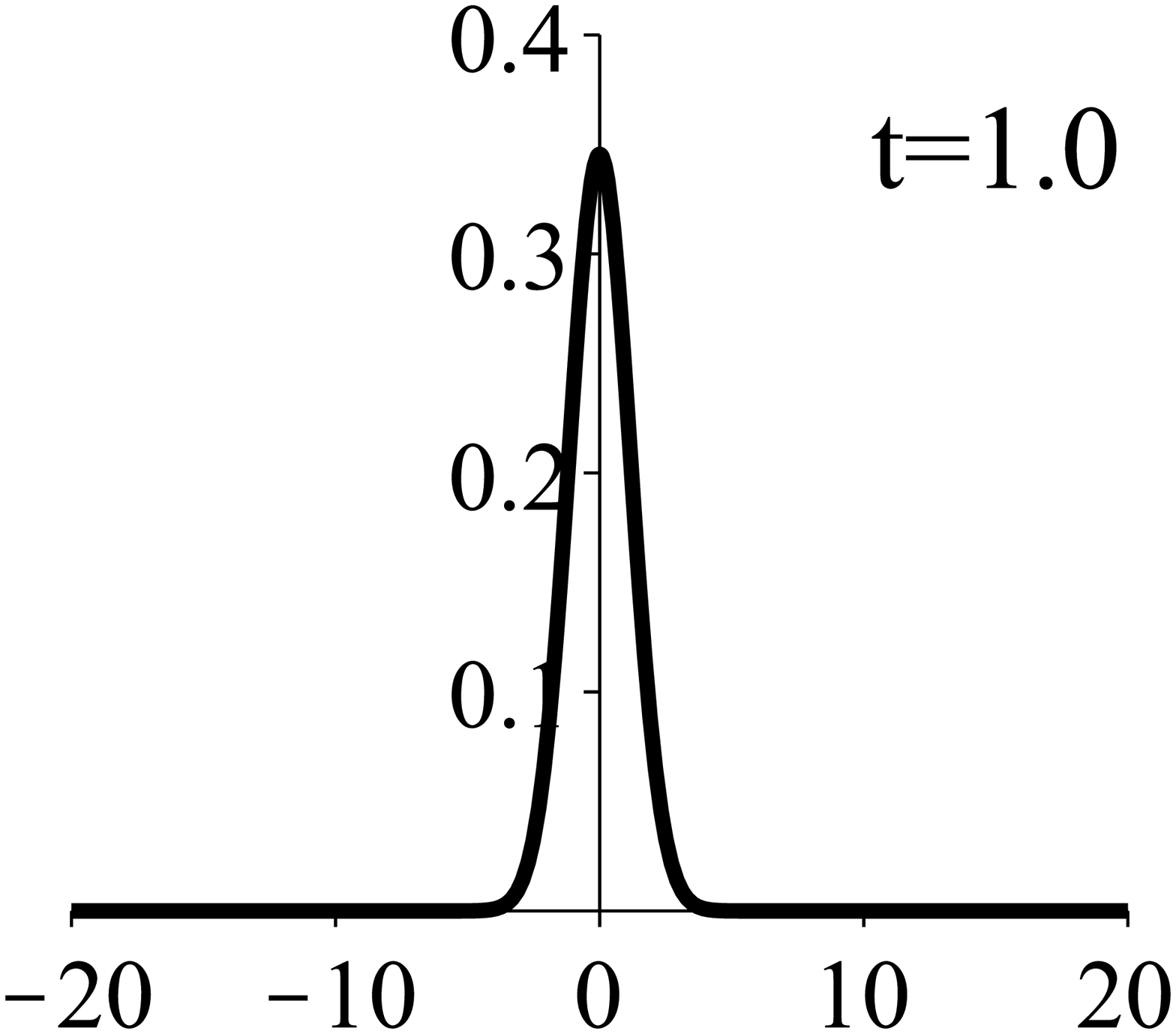}\hspace{3cm}
\includegraphics*[width=3.3cm,height=3.3cm]{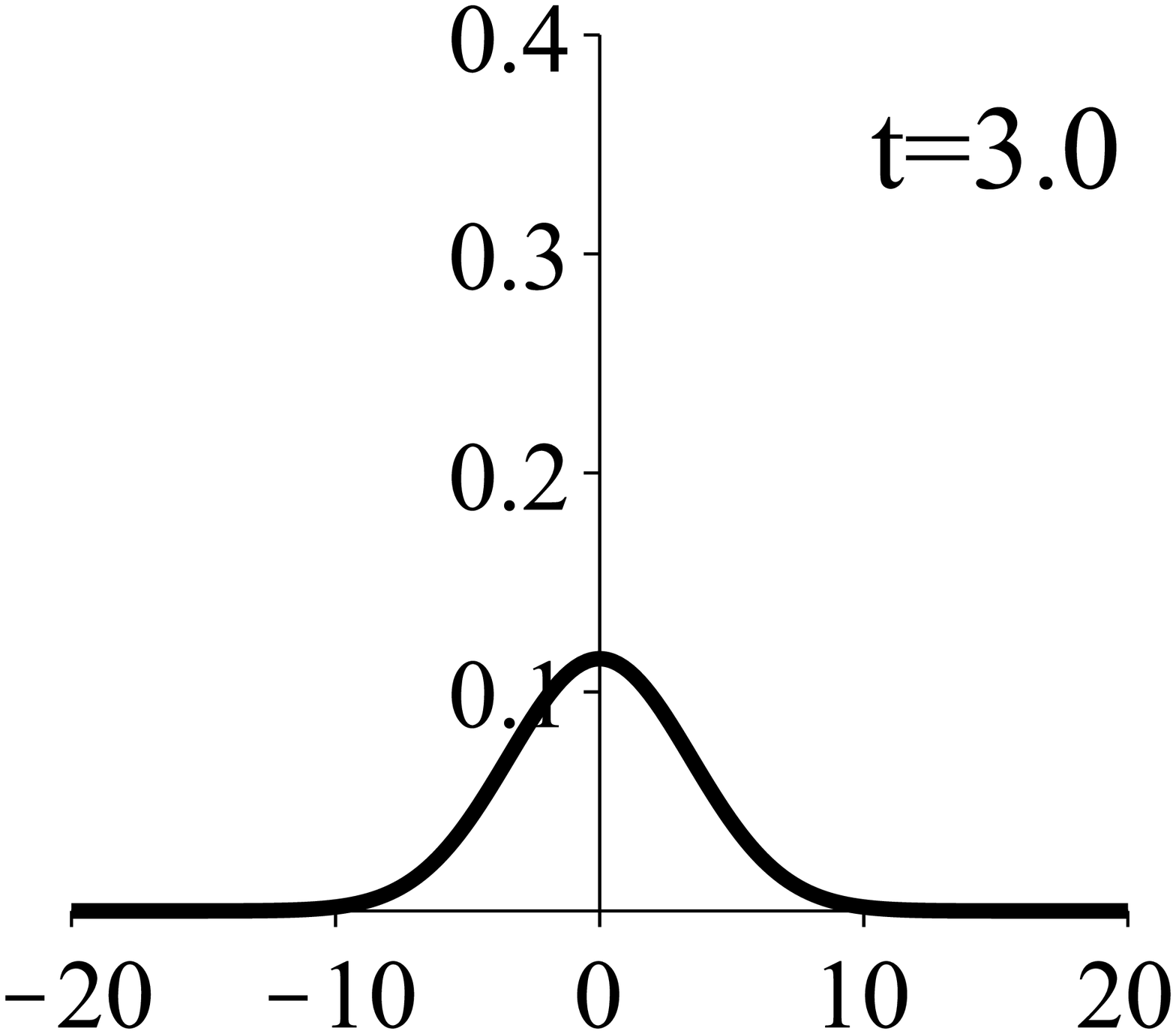}\\
\includegraphics*[width=3.3cm,height=3.3cm]{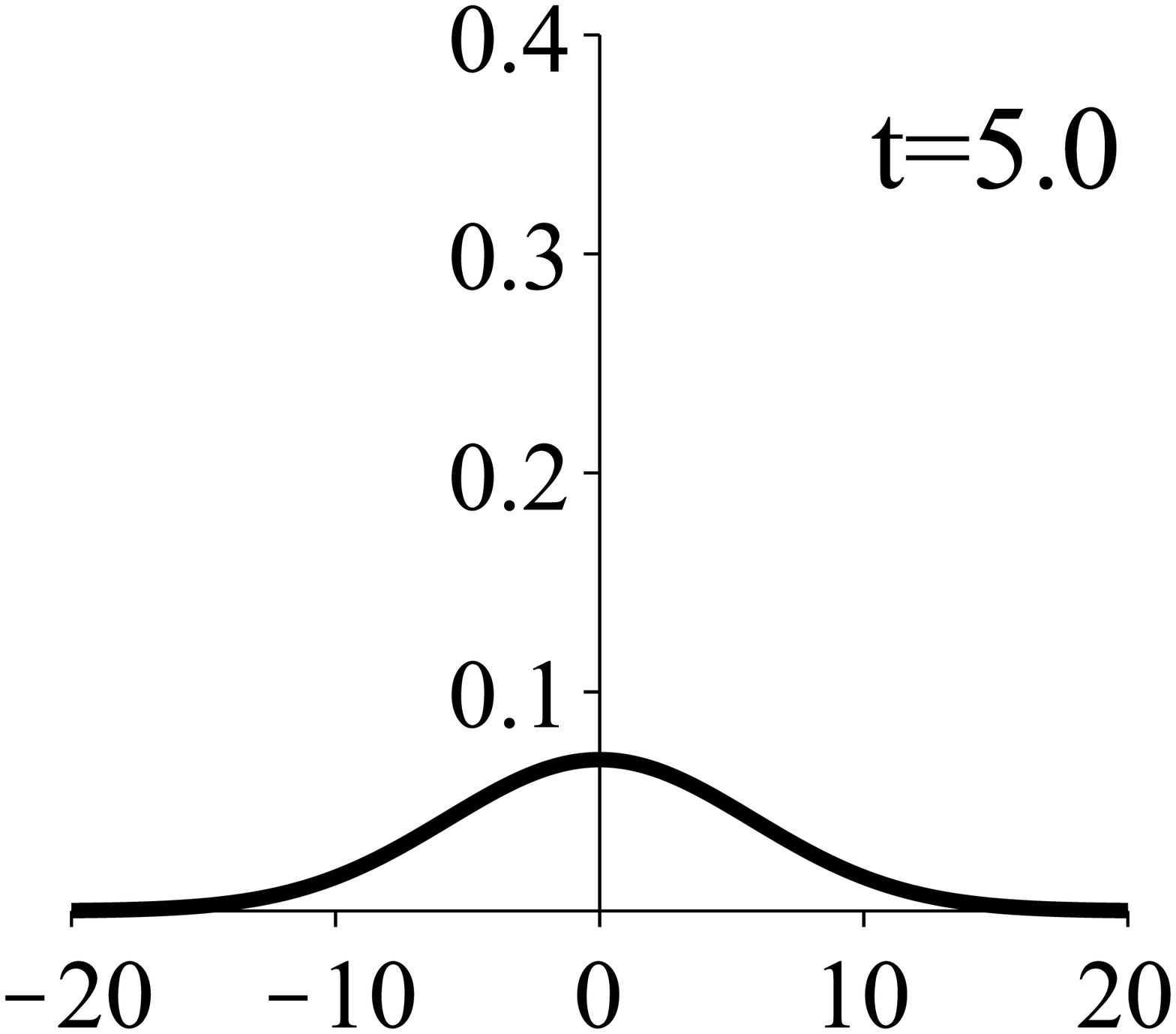}\hspace{3cm}
\includegraphics*[width=3.3cm,height=3.3cm]{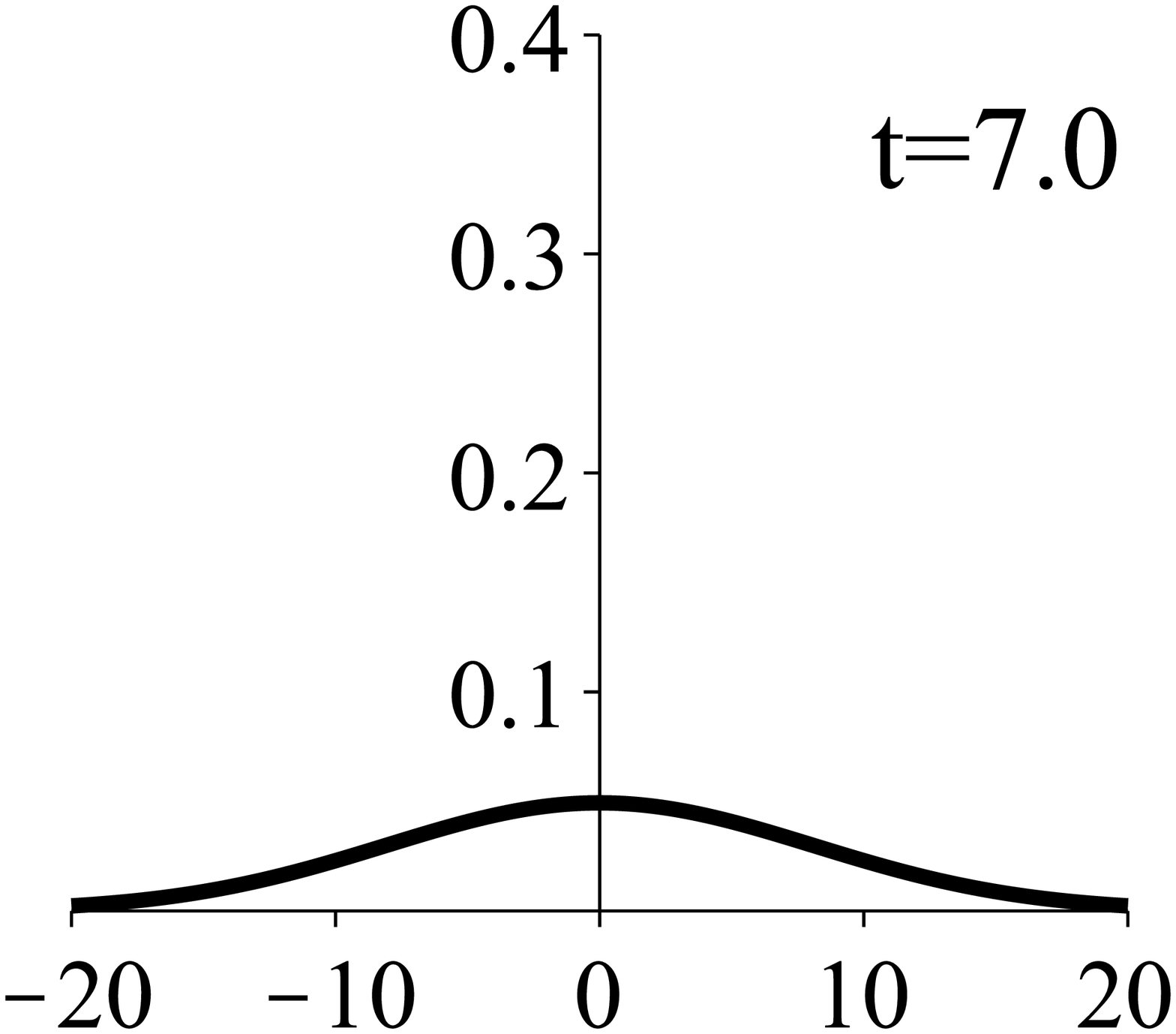}
\caption{Plot of $W(x,t)$ versus $x$ for solution~(\ref{E4.7})
with $\alpha=1$, $\mu_1=1/4$, $\mu_2=0$, $\mu_4=1$, and time
$t=1.0$, $3.0$, $5.0$, $7.0$} \label{fig:EX3}
\end{figure}

The evolution of the solutions~(\ref{E4.6}) and (\ref{E4.7}) are
plotted in FIG.~\ref{fig:EX2} and FIG.~\ref{fig:EX3}
respectively with two sets of the parameters, which are the same
except $\alpha$.

Because of $\mu_2=0$, the peaks in both of the figures do not move
as expected. It can be seen that, before time $t_c=3$, $W(x,t)$ in
FIG.~\ref{fig:EX2} has a larger FWHM and a smaller peak value than
$W(x,t)$ in FIG.~\ref{fig:EX3}, and after $t_c=3$, the situation
reverses. That is because, when $0<t<1$, the diffusion coefficient
$\mu_4$ causes more diffusion than $\mu_4t$ does.  Hence it leads
the peak value of $W(x,t)$ in FIG.~\ref{fig:EX2} to be smaller
than that in FIG.~\ref{fig:EX3}. However, when $t>1$, the
time-dependent diffusion coefficient $\mu_4t$ starts to have more
influence than the constant diffusion coefficient $\mu_4$ does.
When $t=3$, both $W(x,t)$ has the same value for all $x$. After
$t=3$, the peak value of $W(x,t)$ in FIG.~\ref{fig:EX3} is always
smaller than $W(x,t)$ in FIG.~\ref{fig:EX2}.

\subsection{Examples with the same diffusion but different drift
coefficients}

When the set of parameters $\alpha=1/2$, and $\mu_1=\mu_2=0$
is taken, one has $D^{(1)}(x,t)=0$, $D^{(2)}(x,t)=\mu_4$ and
probability distribution
\begin{align}\label{E4.8}
W(x,t)=\sqrt{\frac{1}{4\pi\mu_4 t}}\,
\mbox{exp}\Big\{-\frac{1}{4\mu_4 t}\,x^2\Big\}
\end{align}
The other set of parameters $\alpha=1/2$ and $\mu_1=0$ leads to
$D^{(1)}(x,t)=\mu_2/\sqrt{t}$, $D^{(2)}(x,t)=\mu_4$ and the
probability distribution
\begin{align}\label{E4.9}
W(x,t)=\sqrt{\frac{1}{4\pi\mu_4 t}}\,
\mbox{exp}\Big\{-\frac{1}{4\mu_4 t}\Big(x-2\mu_2\sqrt{t}\Big)^2\Big\}
\end{align}

In this case, (\ref{E4.8}) is the solution of the diffusion
equation with a constant diffusion coefficient $\mu_4$. The
solution~(\ref{E4.8}) demonstrates only the effect of the
diffusion.  Eq.~(\ref{E4.9}) gives the solution with additional
drift $\mu_2/\sqrt{t}$. The probability distribution (\ref{E4.9})
moves with time but keeps its FWHM at the value of $4\sqrt{(\ln
2)\,\mu_4 t\,}$, the same as the solution~(\ref{E4.8}). It is easy
to see in (\ref{E4.9}) that when $\mu_2$ takes the positive
(negative) value, $W(x,t)$ will move in the $+x$ ($-x$) direction
as time elapses, and the peak value $\sqrt{1/(4\pi\mu_4\,t)}$ gets
smaller with time.

\begin{figure}[ht]
\centering
\includegraphics*[width=3.3cm,height=3.3cm]{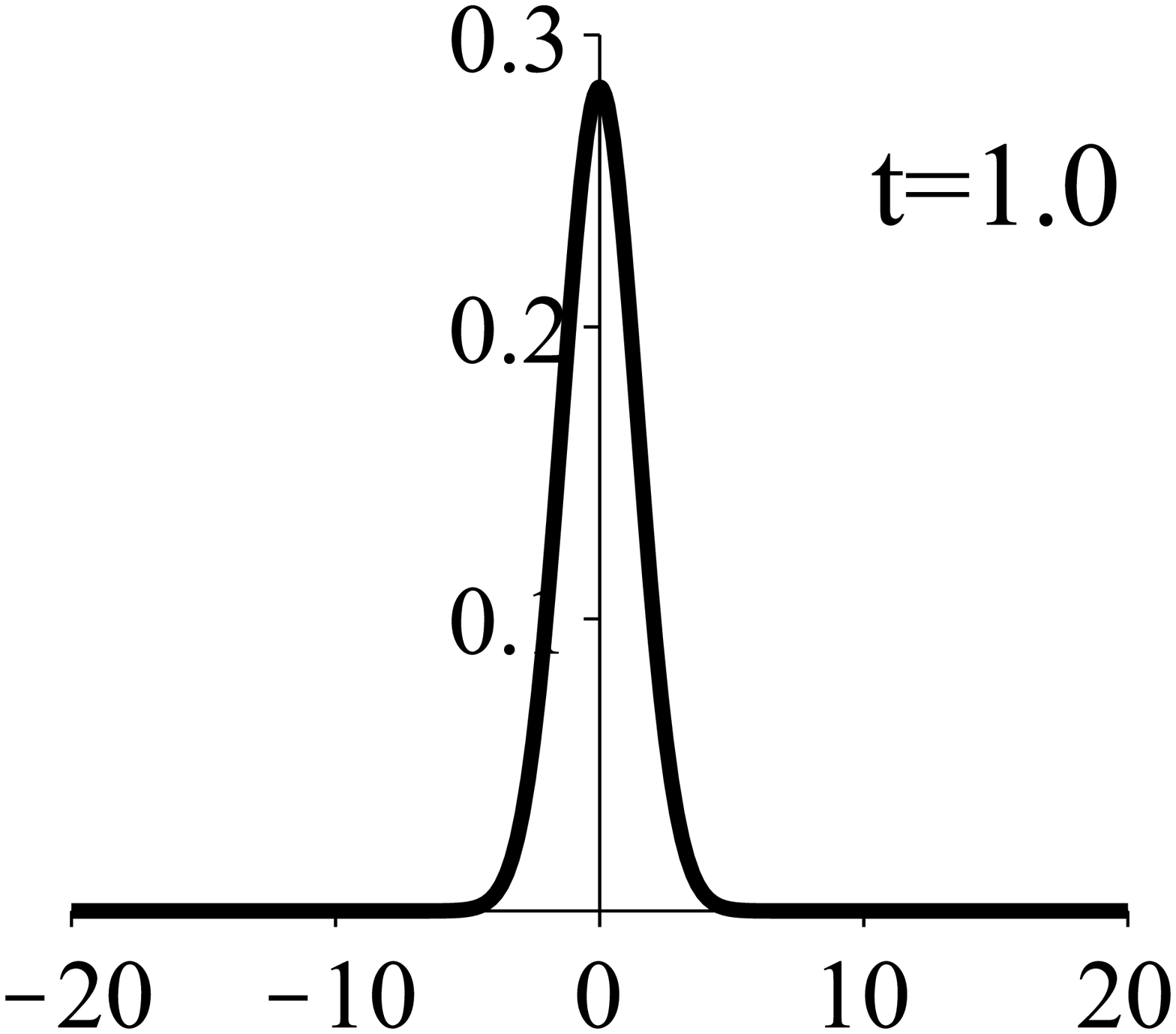}\hspace{3cm}
\includegraphics*[width=3.3cm,height=3.3cm]{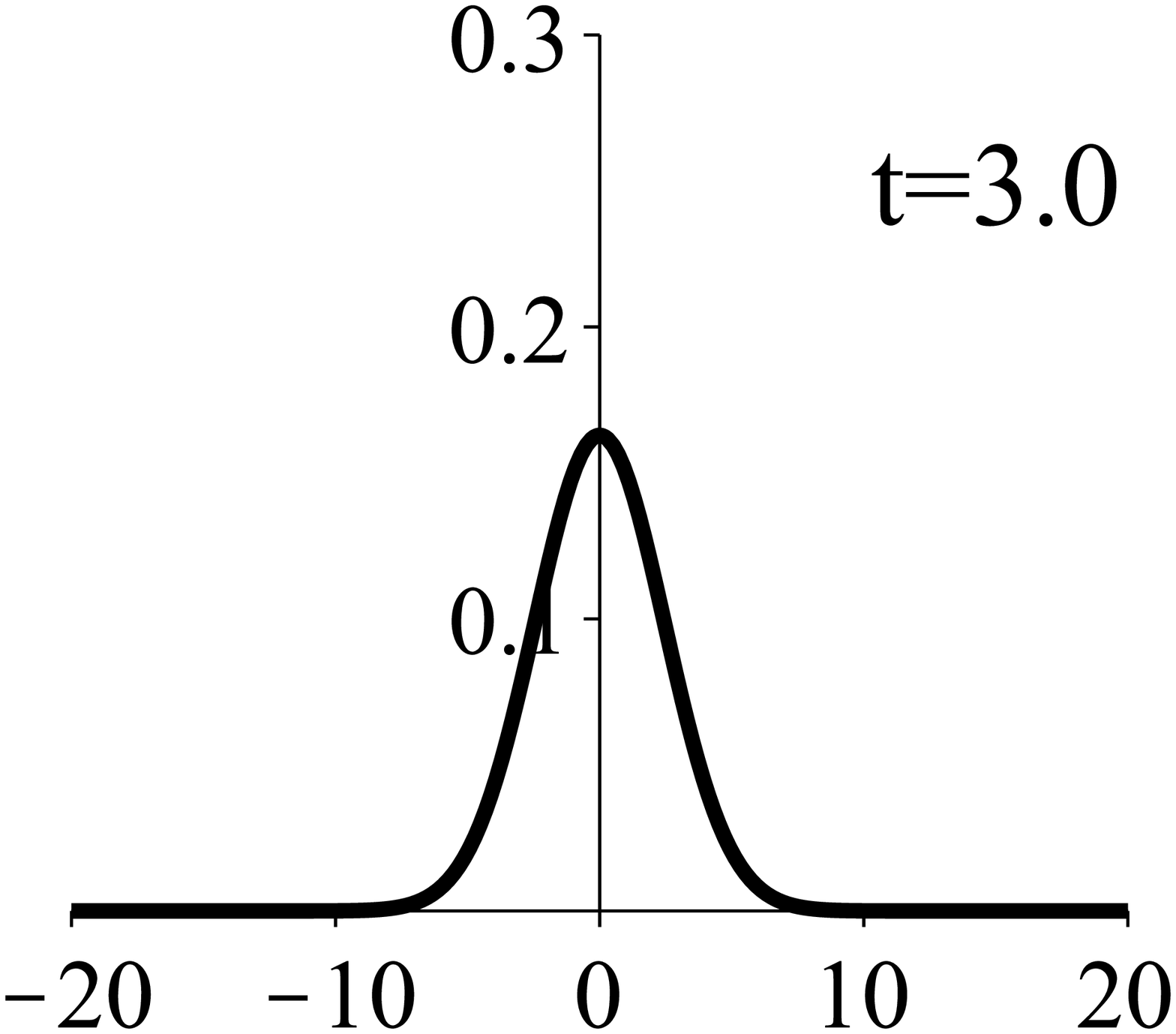}\\
\includegraphics*[width=3.3cm,height=3.3cm]{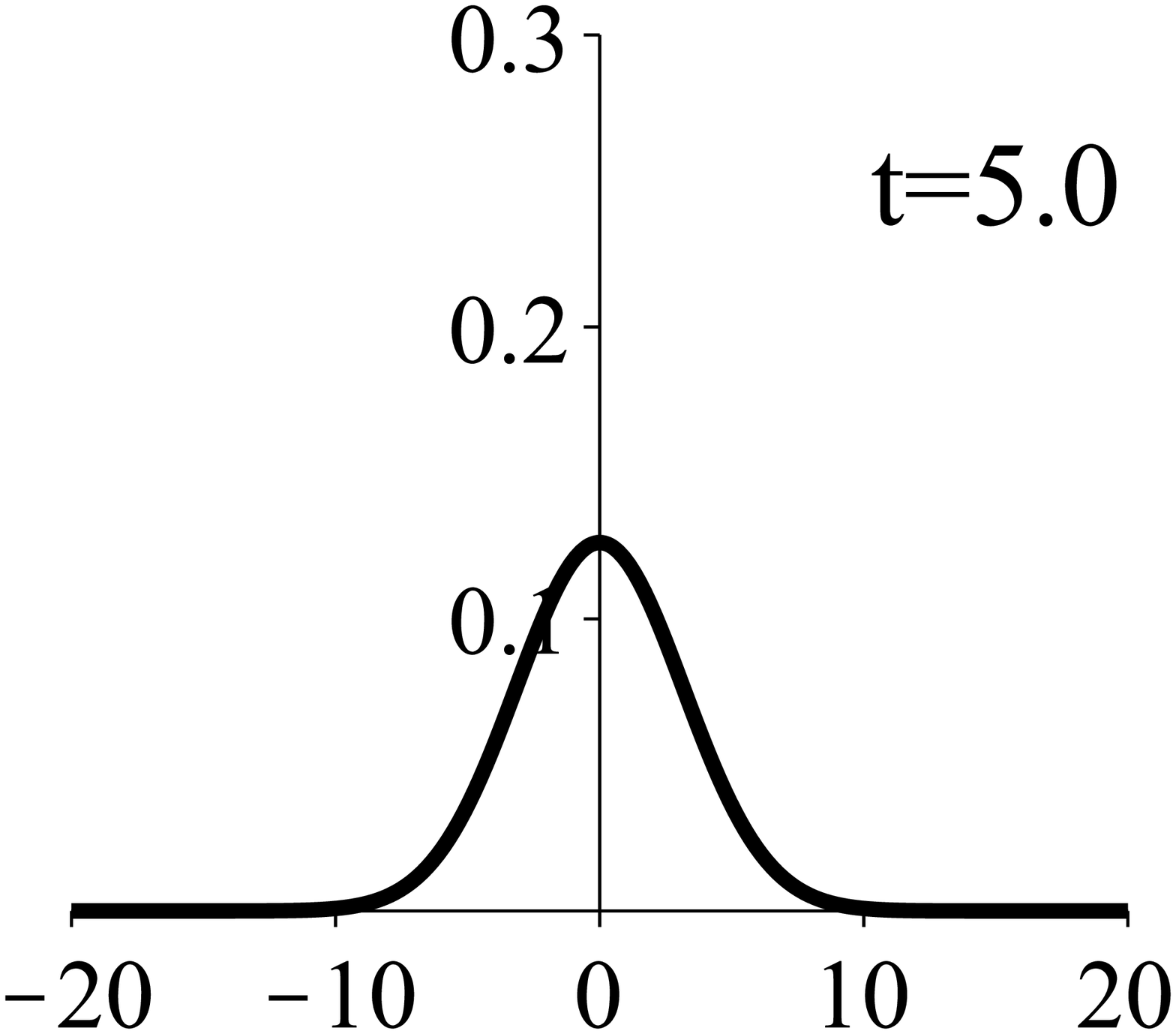}\hspace{3cm}
\includegraphics*[width=3.3cm,height=3.3cm]{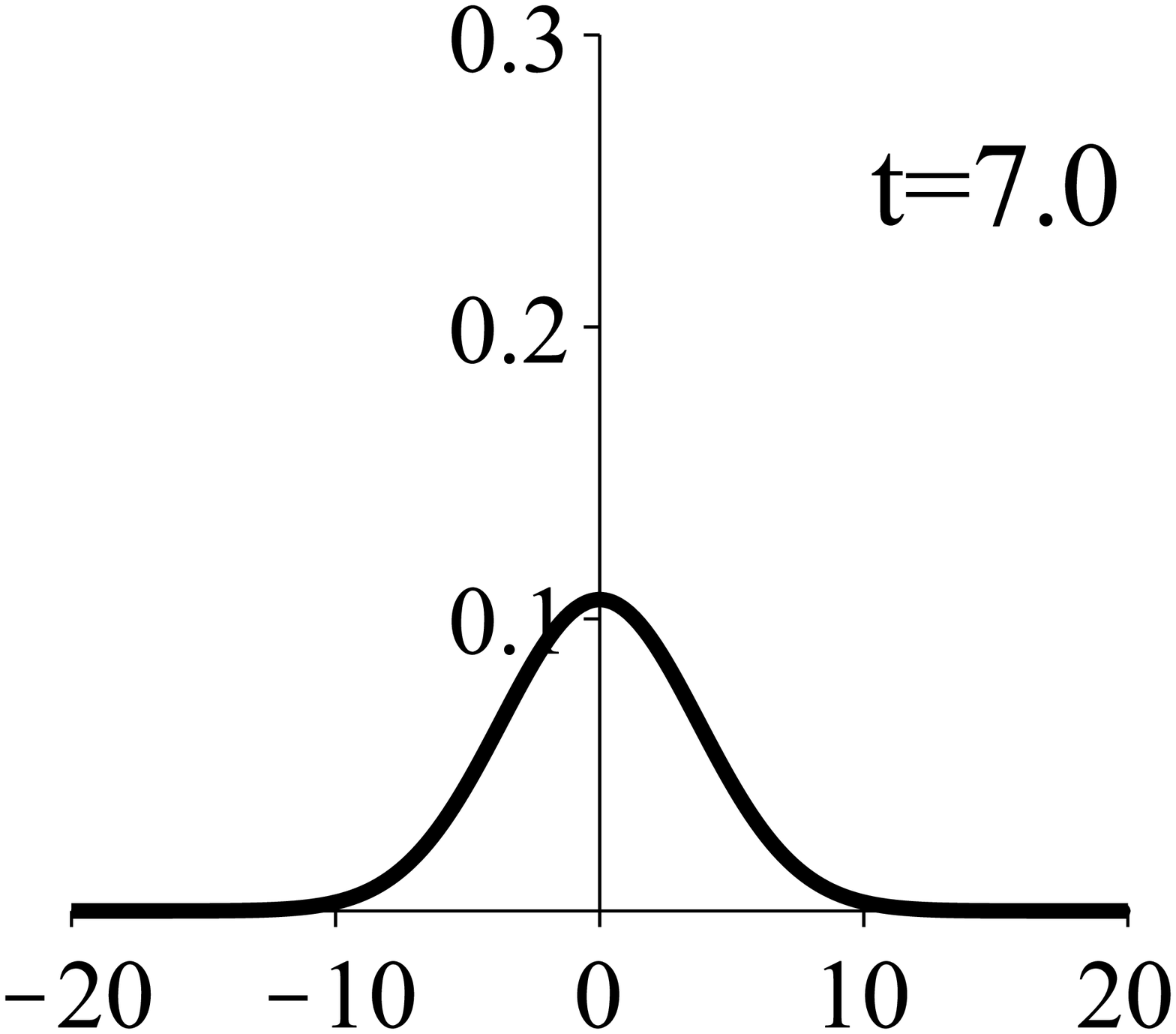}
\caption{Plot of $W(x,t)$ versus $x$ for solution~(\ref{E4.8})
with $\alpha=1/2$, $\mu_1=\mu_2=0$, $\mu_4=1$, and time $t=1.0$,
$3.0$, $5.0$, $7.0$.} \label{fig:EX4}
\end{figure}

\begin{figure}[ht]
\centering
\includegraphics*[width=3.3cm,height=3.3cm]{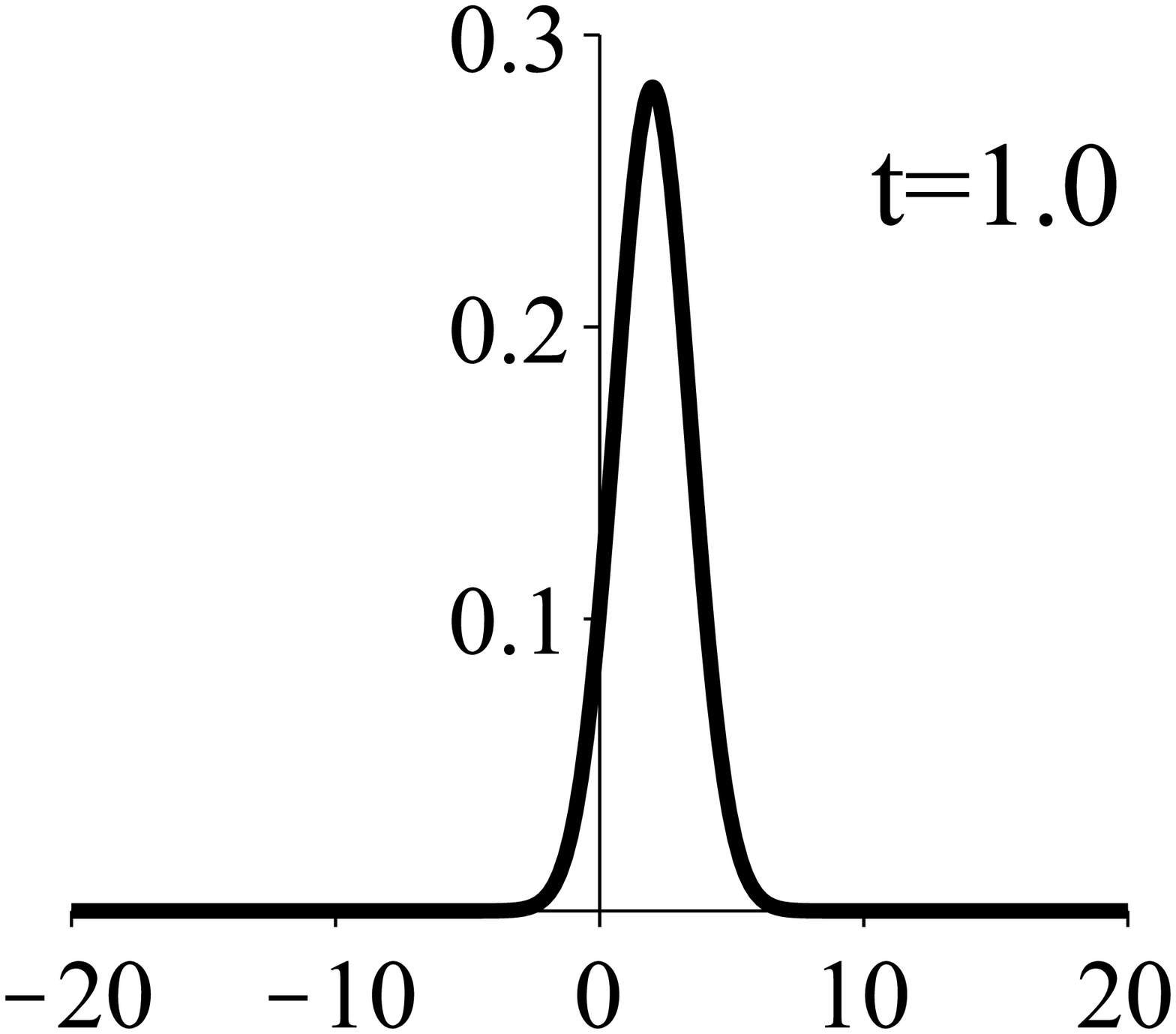}\hspace{3cm}
\includegraphics*[width=3.3cm,height=3.3cm]{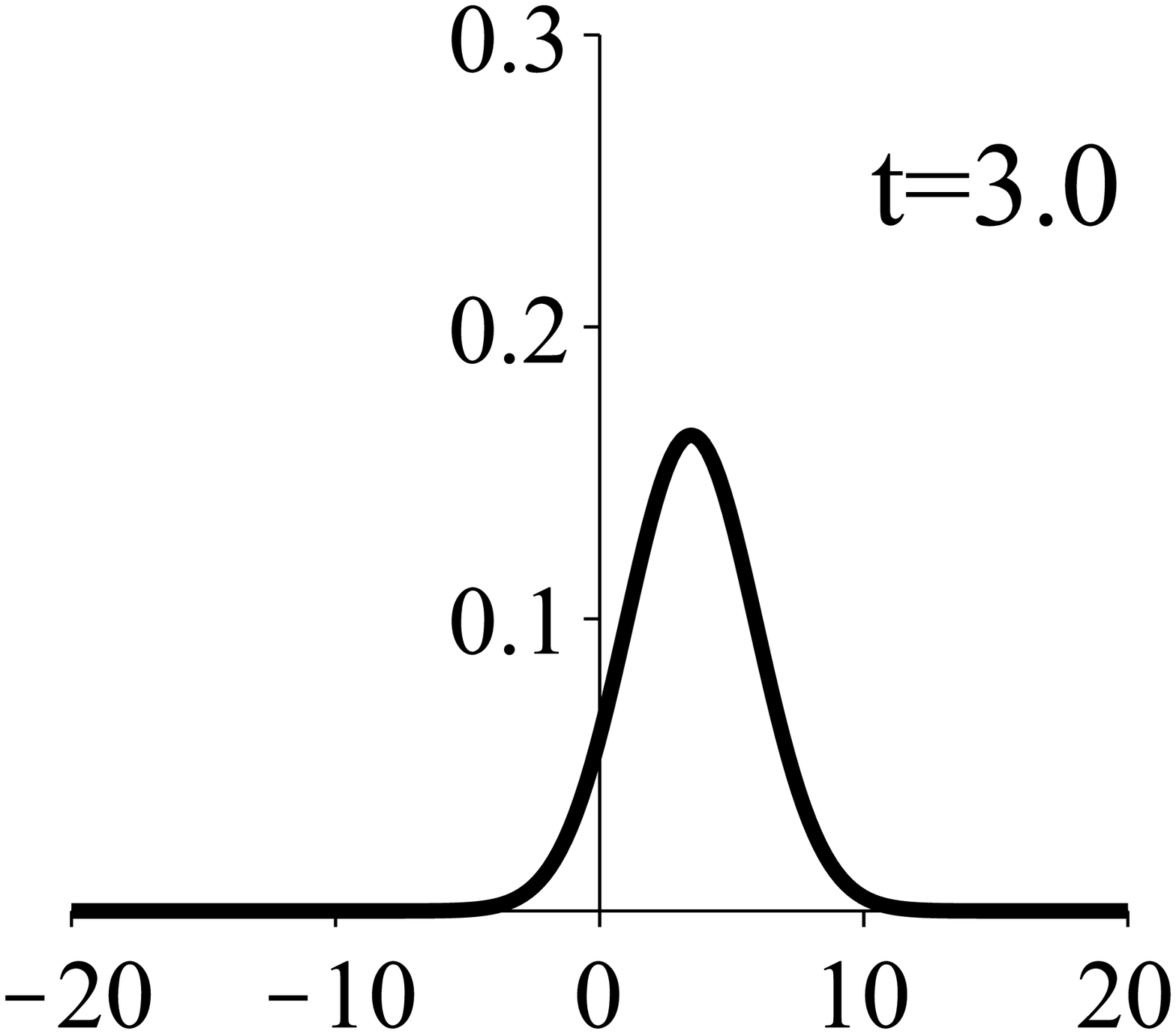}\\
\includegraphics*[width=3.3cm,height=3.3cm]{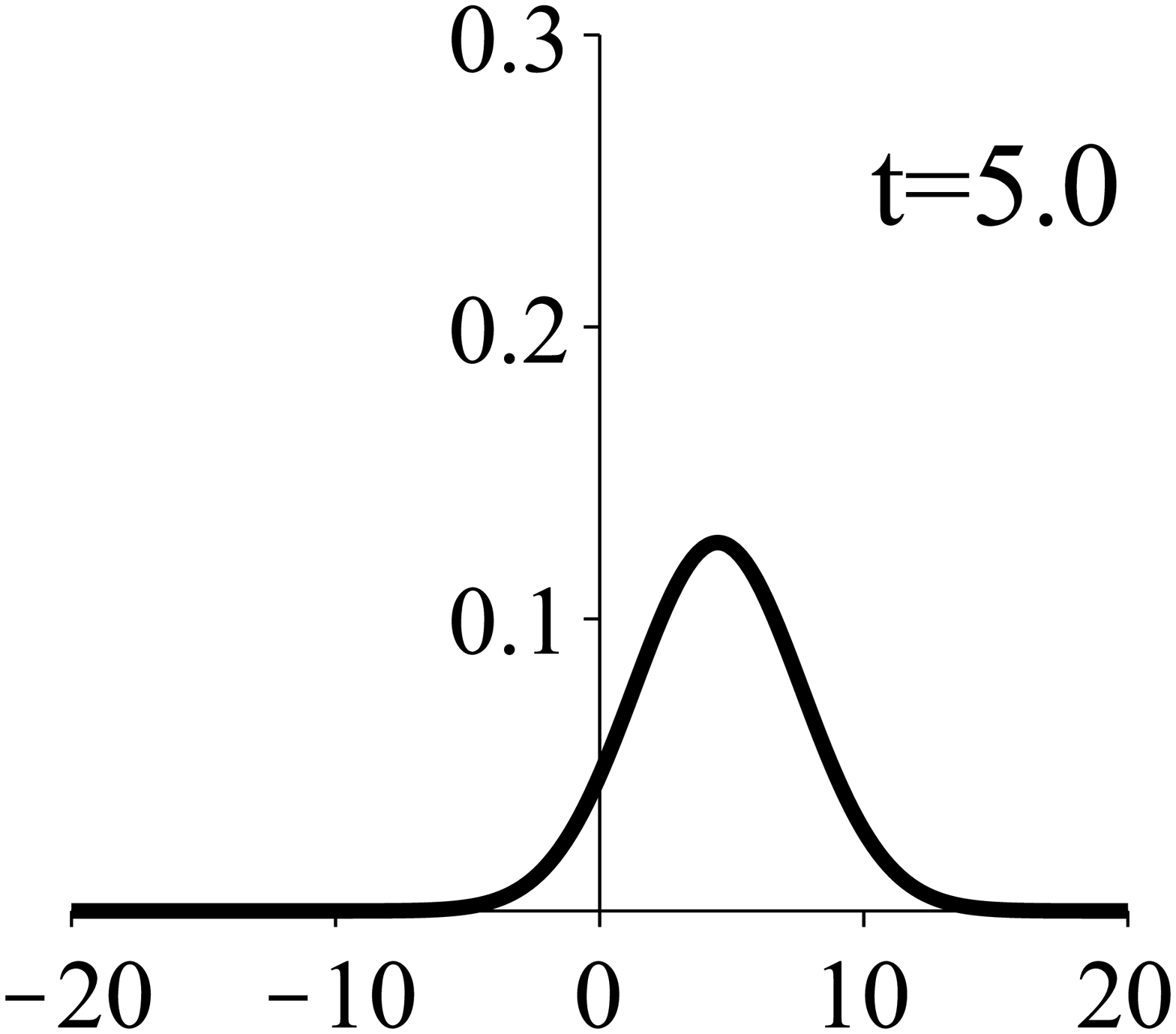}\hspace{3cm}
\includegraphics*[width=3.3cm,height=3.3cm]{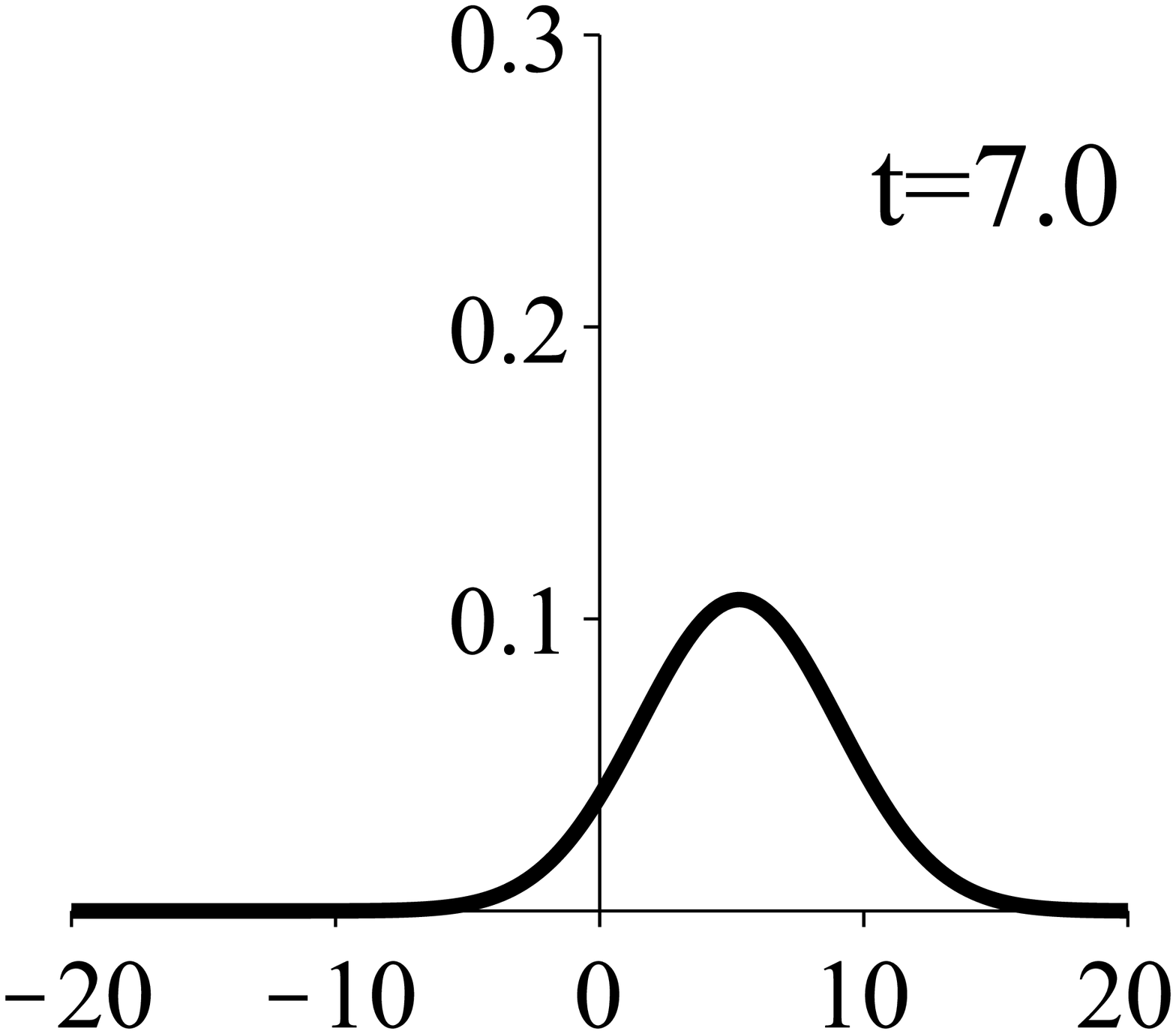}
\caption{Plot of $W(x,t)$ versus $x$ for solution~(\ref{E4.9})
with $\alpha=1/2$, $\mu_1=0$, $\mu_2=1$, $\mu_4=1$, and time
$t=1.0$, $3.0$, $5.0$, $7.0$.} \label{fig:EX5}
\end{figure}

The evolution of these solutions ~(\ref{E4.8}) and (\ref{E4.9})
are plotted in FIG.~\ref{fig:EX4} and FIG.~\ref{fig:EX5}
respectively with two sets of the parameters, which are the same
except $\mu_2$.

\section{FPE with $\rho_1(z)=\mu_1z+\mu_2$ and $\rho_2(z)=\mu_3 z$}

The next interesting exactly solvable example we shall discuss is
an FPE  with $\rho_1(z)=\mu_1z+\mu_2$ and $\rho_2(z)=\mu_3z$. The
corresponding drift and diffusion coefficients are
\begin{align}
D^{(1)}(x,t)=\mu_1\,\frac{x}{t}+\mu_2\,t^{\alpha-1}\;\;\;,\;\;\;
D^{(2)}(x,t)=\mu_3\,x\,t^{\alpha-1}\;.
\end{align}
Eq.~(\ref{W-soln}) is integrable and gives
\begin{align}\label{E4.12}
W(x,t)=\frac{\left|\frac{\alpha-\mu_1}{\mu_3\,t^{\alpha}}\right|^{\frac{\mu_2}{\mu_3}}}
{\Gamma\Big(\displaystyle\frac{\mu_2}{\mu_3}\Big)}\,x^{\frac{\mu_2}{\mu_3}-1}\,
\exp\Big\{-\frac{\alpha-\mu_1}{\mu_3\,t^{\alpha}}\,x\Big\}.
\end{align}
The form of $W(x,t)$ implies that the domain of $x$ is defined
only on half-line.  For definiteness we shall take $x\in
[0,\infty)$.  Normalizability of $W(x,t)$ then requires
\begin{equation}
\frac{\alpha-\mu_1}{\mu_3}>0,~~\frac{\mu_2}{\mu_3}\geq 1.
\end{equation}

It is seen that the distribution $W(x,t)$ is invariant under the
changes $(\mu_2, \mu_3, x)$ and $(-\mu_2, -\mu_3,-x)$ (other
parameters being the same). This means that the distributions
$W(x,t)$ with the parameters $(\mu_2, \mu_3)$ and $(-\mu_2,
-\mu_3)$ are mirror images of each other with respect to the
$y$-axis.  It is also worth noticing that for a given $\alpha$,
one will have the same solutions for different drift and diffusion
coefficients as long as the quantity $(\mu_1-\alpha)/\mu_3$
remains unchanged.  Furthermore, the system is invariant under the
transformations $\alpha\to -\alpha,~ \mu_1\to\mu_1-2\alpha,~t \to
1/t$, and $\mu_2,~\mu_3$ and $\mu_4$ unchanged.  Hence the
distribution with $(\alpha, \mu_1)$ at time $t$ is the same as the
distribution with $(-\alpha, \mu_1-2\alpha)$ at time $1/t$.

\begin{figure}[ht]
\centering
\includegraphics*[width=3.3cm,height=3.3cm]{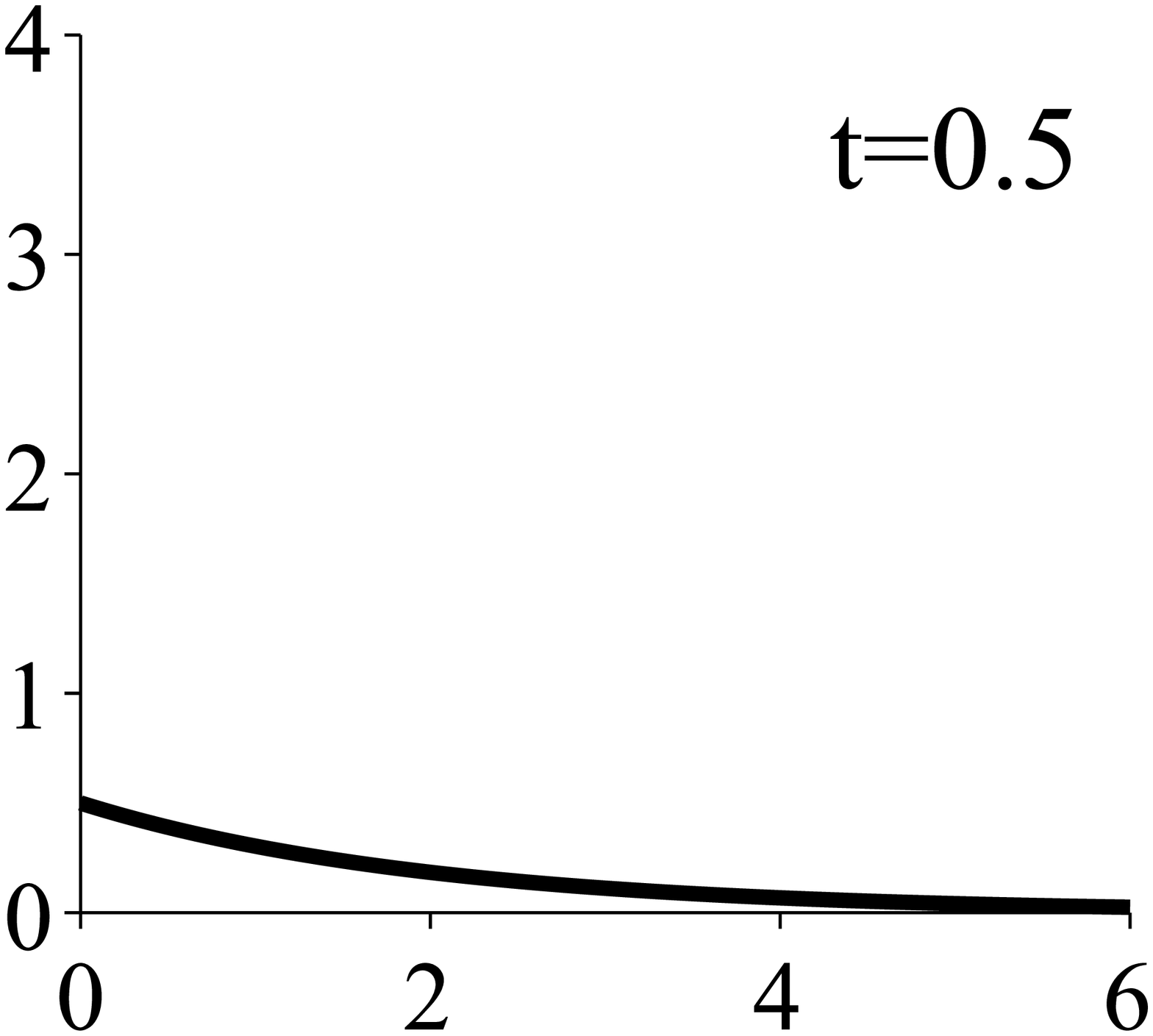}\hspace{3cm}
\includegraphics*[width=3.3cm,height=3.3cm]{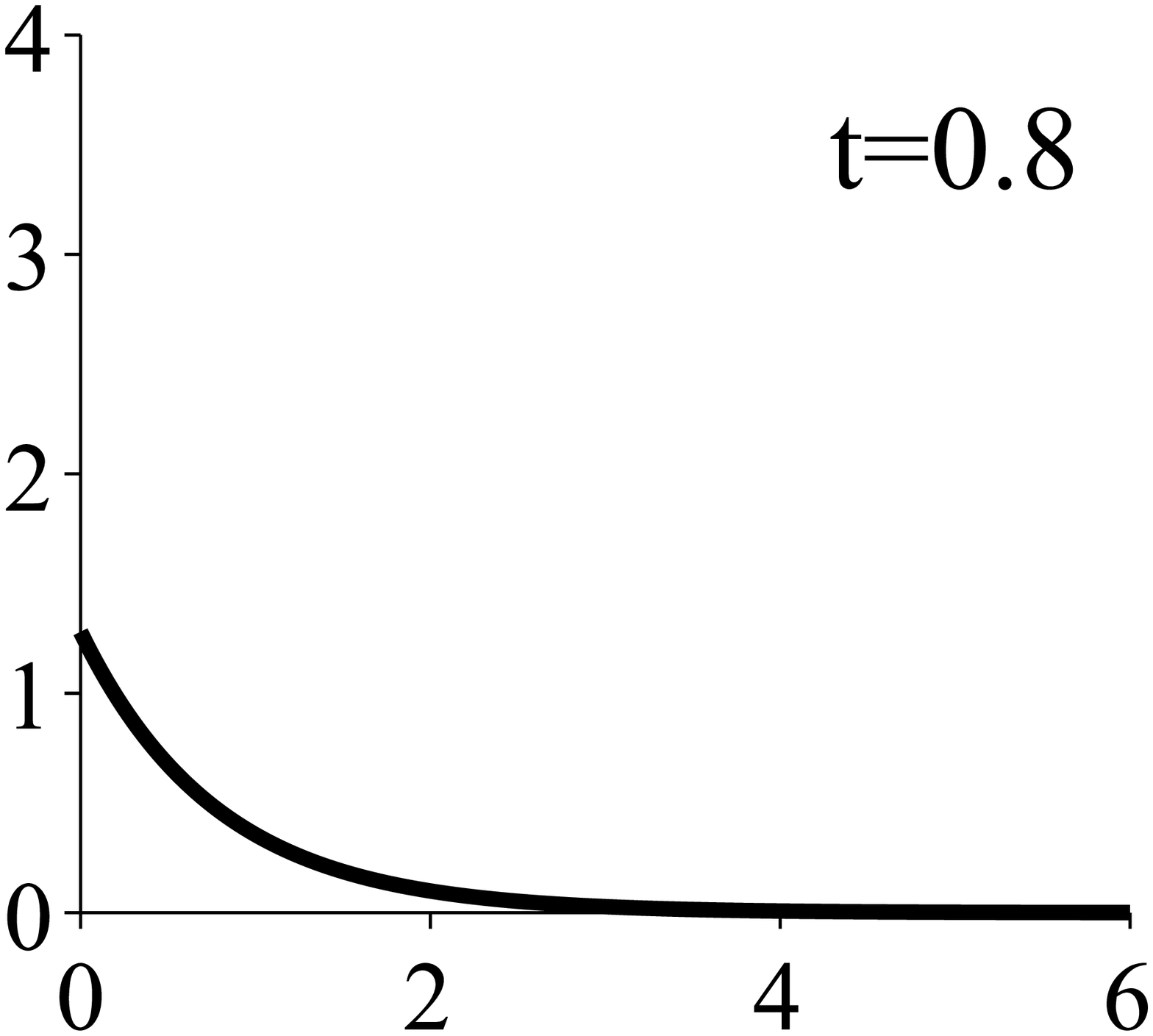}\\
\includegraphics*[width=3.3cm,height=3.3cm]{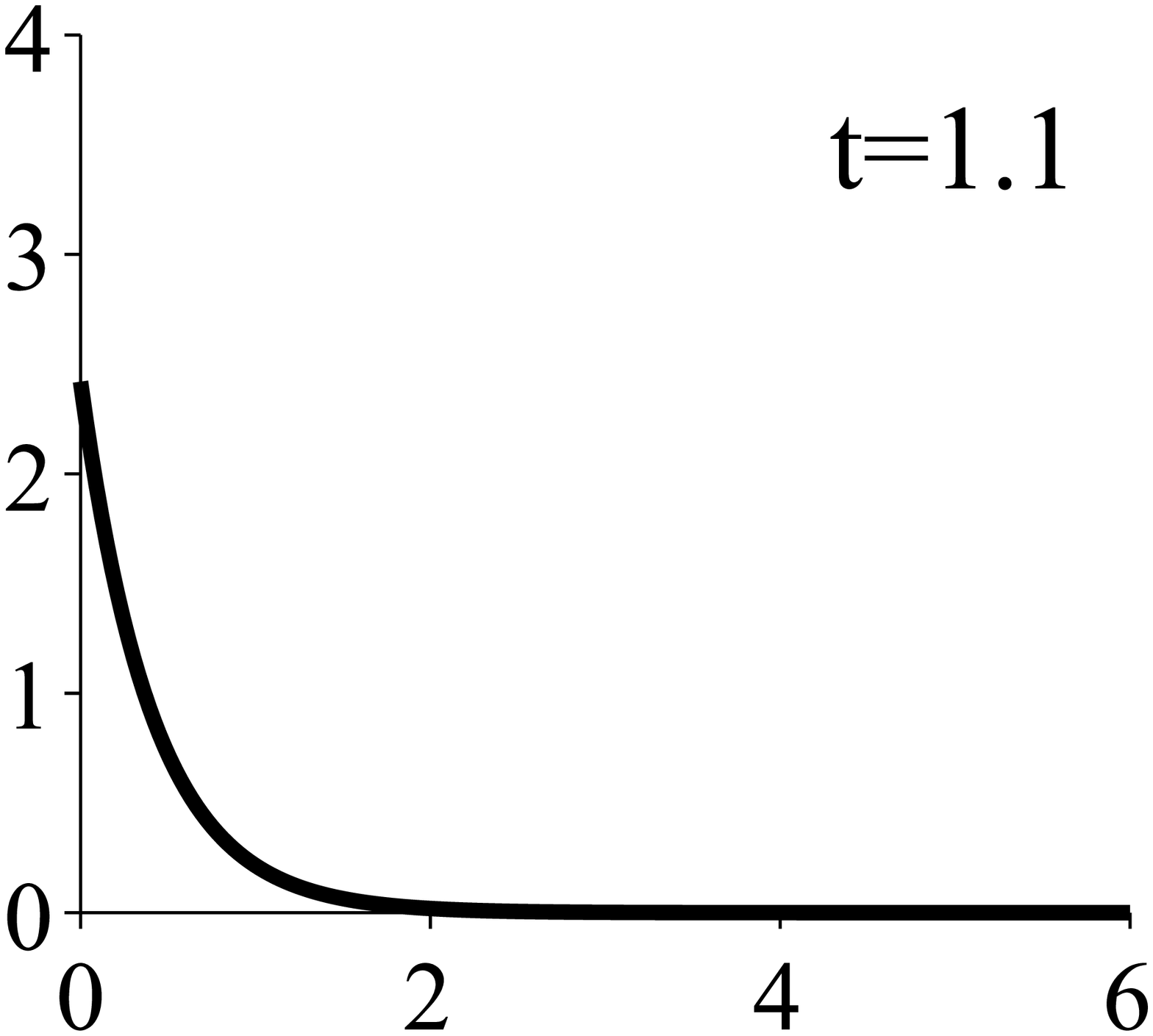}\hspace{3cm}
\includegraphics*[width=3.3cm,height=3.3cm]{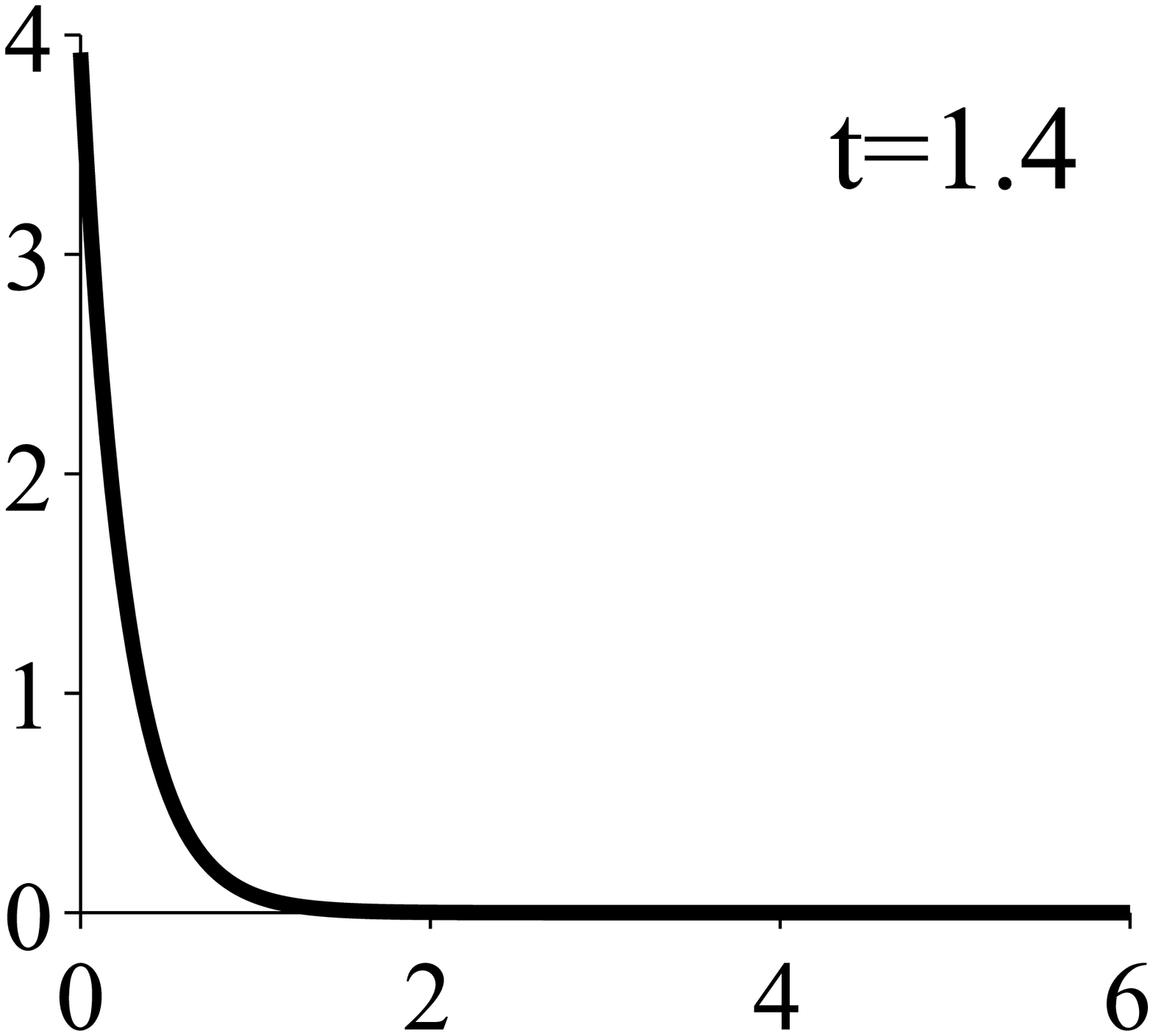}
\caption{Plot of $W(x,t)$ versus $x$ for solution~(\ref{E4.12})
with $\alpha=-2$, $\mu_1=-3$, $\mu_2=\mu_3=1/2$, and time $t=0.5$,
$0.8$, $1.1$, $1.4$.} \label{fig:EX6}
\end{figure}

Solution (\ref{E4.12}) with $\mu_2=\mu_3$ presents an interesting
stochastic process. In this case, $W(x,t)$ becomes the exponential
function, whose peak is always located at the origin. Its peak
value is $|(\mu_1-\alpha)/(\mu_3\,t^{\alpha})|$, which is
dependent on the parameters $\alpha$, $\mu_1$ $\mu_3$ and time
$t$, and hence is affected by both the drift and diffusion
coefficients.  The peak at $x=0$ is increasing (decreasing) as $t$
increases for $\alpha<0$ ($\alpha>0$).  That means, by an
appropriate choice of the drift and diffusion parameters, one can
have a situation where the probability function is accumulating at
the origin. For such situation, the effect of the drift force is
stronger than that of the diffusion,  causing the distribution to
be pushed toward the origin as time elapses. An example of such
situation is depicted in FIG.~\ref{fig:EX6}, which demonstrates
the evolution of solution (\ref{E4.12}) with $\alpha=-2$ and
$\mu_2=\mu_3$.

\newpage

\section{Summary and discussions}

The FPE is an important equation in many different areas. Analytic
solutions of the FPEs with both time-dependent drift and diffusion
coefficients are generally difficult to obtain. In this paper, we
have presented a general way to construct exact similarity
solutions of the FPE. Such similarity solutions exist when the FPE
possesses proper scaling behavior.

The similarity method makes use of the scaling-invariant property
of the FPE.  By the introduction of the similarity variable, the
FPE can be reduced to an ordinary differential equation, which may
be easier to solve. The general expression of the ordinary
differential equation corresponding to the FPE with time-dependent
drift and diffusion coefficients is given in this paper. It is
interesting to find, by the natural requirement that the
probability current density vanishes at the boundary, that the
resulting ordinary differential equation is integrable, and the
probability density function can be given in closed form.  We have
presented several new examples of exactly solvable Fokker-Planck
equations with time-dependent coefficients.  Symmetry properties
of the solutions are also discussed.

Of course there are many other possibilities not presented here.
These can be worked out easily by the reader following the method
presented in this paper.

Now we would like to briefly discuss whether there exist FPEs with
similarity solutions which are related to the so-called
quasi-exactly solvable (QES) equations. QES systems give rise to a
new class of spectral problems for which it is possible to
determine analytically a part of the spectrum but not the whole
spectrum \cite{TU,Tur,GKO,Ush,HR}. Thus QES systems are
intermediate to the exactly solvable systems and the non-solvable
ones. One-dimensional QES systems related to the $sl(2)$ Lie
algebra have been classified into ten classes in \cite{Tur}. For
example, the Class I QES model is given by the equation
\cite{Tur,HR}
\begin{eqnarray}
\mu z^2 y^{\prime\prime}(z)
-\left[2az^2-(2b+\mu)z-2c\right]y^\prime(z)+\left[2aNz+\frac{E}{\mu}\right]y(z)=0,
~~N=0,1,2,\ldots \label{QES-I}
\end{eqnarray}
Note that the parameter $N$ appears in the $z$-dependent term in
the coefficient of $y(z)$.  For each $N=0,1,2,\dots$, there are
only $N+1$ exactly known solutions, and hence it is only QES.

Now one is interested to know whether there exists an FPE whose
similarity solution is related to the $sl(2)$-based QES models. To
answer this question, let us note that if a given second order
differential equation of the form
$P(z)y^{\prime\prime}+Q(z)y^\prime + R(z)y=0$ can be cast in the
form in eq.~(\ref{E3.4}), then one must have
$Q^\prime(z)=P^{\prime\prime}(z)+R(z)$. It is easy to see that
eq.~(\ref{QES-I}) does not satisfy this requirement, and thus
there is no FPE which can be related to the Class I QES model. In
fact, it can be checked that this is also true for the other
classes in \cite{Tur}.

%\newpage

\section*{Acknowledgments}

 This work is supported in part by the
National Science Council (NSC) of the Republic of China under
Grants NSC-99-2112-M-032-002-MY3 and  NSC-99-2811-M-032-012.

\newpage

%\bibliographystyle{plain}
%\begin{thebibliography}{999}

\newpage
%---------   Figures -------------------

%--------   Fig.1 -----------------

%--------   Fig.2  & 3-----------------

%--------   Fig.4 & 5 -----------------

%--------   Fig.6 & 7-----------------

%--------   Fig.8 -----------------


\begin{thebibliography}{55}

\bibitem{RIS:1996}
H. Risken, {\it The Fokker-Planck Equation} (2nd. ed.)
(Springer-Verlag, Berlin, 1996).

\bibitem{BC:1974}
G. W. Bluman and J. D. Cole, {\it Similarity Methods for
Differential Equations} (Springer-Verlag, New York, 1974).

\bibitem{CCR:1981}
B. Caroli, C. Caroli, and B. Roulet, {\it Diffusion in a bistable potential: The functional integral approach},
 J. Stat. Phys. 26, 83 (1981).

\bibitem{DM:1996}
A. N. Drozdov and M. Morillo, {\it Solution of nonlinear Fokker-Planck equations}, Phys. Rev. E 54, 931 (1996).

\bibitem{ZTZ:2005}
O. C. Zienkiewicz, R. L. Taylor and J. Z. Zhu, {\it The Finite Element Method: Its Basis and Fundamentals},
(6th ed.) (Butterworth-Heinemann, 2005)

\bibitem{GMNT:2005}
G. H. Gunaratne, J. L. McCauley, M. Nicol and A. T\"or\"ok, {\it
Variable step random walks and self-similar distributions}, J.
Stat. Phys., Vol.121, No.5-6, 887 (2005).

\bibitem{KSF:2005}
Kwok Sau Fa, {\it Exact solution of the Fokker-Planck equation for a broad class of diffusion coefficients},
Phys. Rev. E 72, 020101(R) (2005).

\bibitem{GNT:2009}
G. H. Gunaratne, M. Nicol and A.
T$\ddot{\mbox{o}}$r$\ddot{\mbox{o}}$k, {\it Clustering of
volatility in variable diffusion processes}, Physica A 388, 4424
(2009).

\bibitem{LM:2000}
F. Lillo and R. N. Mantegna, {\it Drift-controlled anomalous diffusion: A solvable Gaussian model},
Phys. Rev. E 61, R4675 (2000).

\bibitem{HY:2008}
C.-L. Ho and Y.-M. Dai, {\it A perturbative approach to a class of
Fokker-Planck equations}, Mod. Phys. Lett. B 22, 475 (2008).

\bibitem{LH:2011}
W.-T. Lin and C.-L. Ho, {\it Similarity solutions of a class of
perturbative Fokker-Planck equations}, J. Math. Phys. 52, 073701
(2011).

\bibitem{WH:1980}
W. Weidlich and G. Haag, {\it Quasiadiabatic solutions of Fokker
Planck equations with time-dependent drift and fluctuations
coefficients}, Z. Phys. B 39, 81 (1980).

\bibitem{OK:1985}
J. Owedyk and A. Kociszewski, {\it On the Fokker-Planck equation
with time-dependent drift and diffusion coefficients and its
exponential solutions}, Z. Phys. B 59, 69 (1985).

\bibitem{SS:1999}
S. Spichak and V. Stognii, {\it Symmetry classification and exact
solutions of the one-dimensional Fokker-Planck equation with
arbitrary coefficients of drift and diffusion}, J. Phys. A 32,
8341 (1999).

\bibitem{ZIL:1992}
D. Zwillinger, {\it Handbook of Differential Equations} (2nd. ed.) (Academic Press, 1992).

\bibitem{TU} A. Turbiner and A.G. Ushveridze, {\it Spectral singularities
and quasi-exactly solvable quantal problem}, Phys. Lett. A 126,
181 (1987).

\bibitem{Tur} A.V.Turbiner, {\it Quasi-exactly solvable problems and $sl(2)$ algebra},
Comm. Math. Phys. 118, 467 (1988).

\bibitem{GKO} A. Gonz\'alez, N. Kamran and P.J. Olver, {\it Normalizability of
one-dimensional quasi-exactly solvable Schr\"odinger operators},
Comm. Math. Phys. 153, 117 (1993).

\bibitem{Ush} A.G. Ushveridze,  {\it Quasi-exactly Solvable Models in Quantum
Mechanics} (IOP, Bristol, 1994).

\bibitem{HR} C.-L. Ho and P. Roy, {\it Quasi-exact solvability of
the Pauli equation}, J. Phys. A 36, 4617 (2003).

\end{thebibliography}
\end{document}